\documentclass[conference]{IEEEtran}
\IEEEoverridecommandlockouts

\usepackage{cite}
\usepackage{amsmath,amssymb,amsfonts}
\usepackage{graphicx}
\usepackage{textcomp}
\usepackage{xcolor}
\usepackage{booktabs}
\usepackage{array}
\usepackage{graphicx}
\usepackage{multirow}
\usepackage{tabularx}
\usepackage{textcomp}
\usepackage[ruled,vlined,linesnumbered]{algorithm2e}
\usepackage{algorithmic}
\usepackage{ragged2e}
\usepackage{siunitx}
\graphicspath{{images/}}

\def\BibTeX{{\rm B\kern-.05em{\sc i\kern-.025em b}\kern-.08em
    T\kern-.1667em\lower.7ex\hbox{E}\kern-.125emX}}

\usepackage{pgfplots}
\usetikzlibrary{arrows.meta, positioning, shapes.geometric, fit, backgrounds}
\usepgfplotslibrary{groupplots}
\pgfplotsset{compat=1.18}
\usepackage{pgfplots}
\usepgfplotslibrary{groupplots,statistics}
\pgfplotsset{compat=1.18}

\definecolor{oiBlue}{HTML}{0072B2}
\definecolor{oiVermillion}{HTML}{D55E00}
\definecolor{lightGrid}{HTML}{D9D9D9}
\definecolor{segGray}{HTML}{8C8C8C}
\definecolor{colBL}    {RGB}{ 70,130,180}
\definecolor{colBLPID} {RGB}{ 30, 70,140}
\definecolor{colBLD}   {RGB}{210, 80, 45}
\definecolor{colBLDPID}{RGB}{220,170,  0}

\tikzset{
    basepoint/.style={
        only marks,
        mark=*,
        mark size=1.45pt,
        draw=oiBlue,
        fill=oiBlue
    },
    pidpoint/.style={
        only marks,
        mark=square*,
        mark size=1.35pt,
        draw=oiVermillion,
        fill=oiVermillion
    },
    dumbbellseg/.style={
        segGray,
        line width=0.55pt,
        opacity=0.85
    }
}

\begin{document}

\title{Lightweight PID-Based Drift Mitigation for Cellular Traffic Forecasting}

\author{
\IEEEauthorblockN{
John Sengendo\textsuperscript{1},
Zineddine Bettouche\textsuperscript{2},
Khalid Ali\textsuperscript{2},
Andreas Kassler\textsuperscript{2},
Fabrizio Granelli\textsuperscript{1}
}
\IEEEauthorblockA{
\textsuperscript{1}C.N.I.T and University of Trento, Italy \\
\textsuperscript{2}Deggendorf Institute of Technology, Germany \\
Emails: \{john.sengendo, fabrizio.granelli\}@unitn.it, 
\{zineddine.bettouche, khalid.ali, andreas.kassler\}@th-deg.de
}
}

\maketitle

\begin{abstract}
As mobile networks transition from Beyond 5G (B5G) towards 6G, accurate traffic forecasting is a prerequisite for improving network management. However, with increasing heterogeneity and a massive surge in connected devices, combined with dynamically evolving traffic 
patterns, accurate forecasting is a persistent bottleneck. Existing frameworks, while generally effective, often lack efficiency and degrade under drift, thus requiring costly model retraining to restore performance. In this paper, we propose a lightweight error correction framework that improves forecasting accuracy by integrating a Proportional-Integral-Derivative (PID) controller as a correction layer enhancing Hierarchical Spatio-temporal Models 
(HiSTM). Unlike retraining-based model adaptation, our framework performs online error correction without modifying the model parameters. Results from the proposed framework, evaluated across drift scenarios and cell-level analysis, demonstrate reduced Mean Absolute Error (MAE) and Root Mean Squared Error (RMSE), achieving an average drift mitigation of up to 30.18\% in MAE and 26.68\% 
in RMSE, thereby validating the robustness of the PID framework as a drift mitigation mechanism for network traffic forecasting.
\end{abstract}


\section{Introduction}
\label{sec:introduction}

Accurate network traffic forecasting is a fundamental for autonomous 5G and next-generation (6G) network management \cite{Coronado}, especially as cellular device connections continue to grow. With a projection of upto 7.8 billion IoT connections by 2031 \cite{ericsson_iot_outlook}, network operators and internet service providers (ISPs) must proactively allocate resources, avoid congestion, and maintain stringent Quality of Service (QoS) requirements as latest studies underscore \cite{polverini2026avoiding, wang2024survey}. In this context, the ability to accurately forecast spatiotemporal traffic dynamics is essential to enable fully autonomous and efficient network management, especially when the state of a base station (BS) is influenced not only by its own history, but also by neighboring cells \cite{qiu2018spatio}. 

Despite the predictive power of current state-of-the-art frameworks such as deep learning models, they carry significant computational overhead. Approaches ranging from 
Long Short-Term Memory (LSTM) networks and Convolutional Neural Network (CNNs) to hybrid CNN-LSTM and multi-source spatiotemporal architectures \cite{santos20205g, hussien2025machine, althamary2024enhanced} 
have demonstrated strong performance, yet their training and continual retraining demands consume substantial energy and operational costs \cite{liang2024energy, mahadevan2023cost}. This creates a critical efficiency bottleneck for network providers who must simultaneously minimize expenditure and uphold QoS guarantees.

Compounding this challenge is the inherent non-stationarity of network traffic. Temporal and spatial distribution deviations termed concept drift progressively degrade model accuracy as real-world traffic patterns diverge from those seen during training \cite{d2019survey}. Additionally, in the current 5G network environment, drift is not an exception but a persistent operational reality, driven by mobility patterns, load fluctuations, and evolving user patterns. The classical approach to drift is periodic or continuous model 
retraining \cite{gudepu2024generative}. However, retraining large-scale 
spatiotemporal models is computationally expensive, energy-intensive 
\cite{liang2024energy}, and disruptive to real-time operations directly conflicting with the efficiency and latency requirements of next-generation 
networks \cite{zanotti2025global, mahadevan2023cost}. Despite recent advances in spatiotemporal forecasting, existing frameworks depend on this retraining 
cycle to maintain performance under drift, limiting their practicality for real-time and energy-efficient 5G/6G deployments.

A promising approach is found in lightweight feedback-driven correction mechanisms that operate on top of a trained base model. The Proportional–Integral–Derivative (PID) controller, a feedback technique offers this benefit. As a computationally efficient correction layer that continuously adjusts predictions based on observed 
forecasting error, it removes the need for model retraining and modifying any model weights. By treating cumulative error as an integral signal and responding to rapid error changes through its derivative term, a PID controller can dynamically compensate for drift, offering a principled and resource-efficient complement to traffic forecasting models.
The key contributions of this article are as follows:

\begin{itemize}
    \item We propose a PID-based correction framework that integrates baseline spatiotemporal forecasting models to enhance forecasting accuracy without model retraining.
    
    \item To ensure realistic evaluation, our framework is validated on a real-world 5G network spatiotemporal traffic dataset.
    
    \item We conduct a systematic drift analysis by injecting concept drift into the traffic data, demonstrating that the proposed PID mechanism 
    enhances adaptability and robustness under dynamic network conditions.
    
    \item We perform fine-grained per-cell PID parameter tuning, demonstrating improved prediction and drift mitigation at the individual spatial cell level.
\end{itemize}

The remainder of the paper is organized as follows. 
Section~\ref{sec:related} reviews related works. 
Section~\ref{sec:methodology} describes the methodology. 
Section~\ref{sec:experimental} presents the experimental work flow and evaluation. Section~\ref{sec:analysis} discusses results obtained, while section~\ref{sec:conclusion} concludes the article and outlines future works.
\section{Related Work}
\label{sec:related}
In this section, we review and discuss the current state-of-the-art literature in line with forecasting, drift and residual correction.

\subsection{Statistical Forecasting Mechanisms}
As literature underscores \cite{kochetkova2023short}, early network traffic forecasting relied on statistical time-series models 
such as Autoregressive Integrated Moving Average (ARIMA), seasonal ARIMA, Holt-Winters, and Kalman filtering, valued 
for their interpretability and low computational cost \cite{kochetkova2023short, 
han2021abstracted, moreno2024intelligent, mosahebfard2024intelligent}. 
While these frameworks demonstrated effectiveness for short-term stationary forecasting tasks, their assumptions of 
linearity and stationarity limits their capacity to capture the nonlinear, spatiotemporal, and heterogeneous behavior characteristic of large-scale 5G/6G network traffic, thus necessitating the adoption of more advanced mechanisms.

\subsection{Deep Learning and Spatiotemporal Models}
Recurrent neural networks (RNNs) and Long Short-Term Memory (LSTM) models have shown improved temporal dependency 
modeling and forecasting, while CNN-RNN, CNN-LSTM, transformer-based, and multi-source architectures further advanced cellular traffic forecasting as earlier studies in \cite{azzouni2017long, hussien2025machine} and \cite{althamary2024enhanced} have demonstrated. However, many sequence models treat spatial regions independently, overlooking correlated traffic dynamics across neighboring cells driven by mobility and dynamic local demands \cite{ngo2024flexible}. Graph Neural Networks (GNNs) and spatiotemporal frameworks address this by jointly modeling spatial dependencies and temporal evolution. GNN-based 
models exploit topological and correlation structures among cells, demonstrating strong performance on 5G beam-level traffic forecasting \cite{tommy2025spatio, wang2022spatial, song2024spatio, patidar2025spatio}. Within the same line of spatiotemporal traffic, the Hierarchical Spatiotemporal Mamba (HiSTM) discussed by the latest works in \cite{bettouche2025histm} combines stacked spatial convolutional encoders, a Mamba-based temporal module, and attention-based aggregation to efficiently capture long-range spatiotemporal dependencies with strong generalization and computational efficiency \cite{bettouche2025histm}.

\subsection{Drift Adaptation in Cellular Forecasting}
Mobile traffic is inherently non-stationary. Mobility patterns, service demand shifts, special events that attract crowds, and infrastructure changes can introduce 
concept drift or covariate shift, where deployment data diverges from training distributions \cite{Gama, nugraha2025novel, tziouvaras2025towards, 
singh2013quantifying}. In cellular networks, drift may be spatially localized affecting specific cells or regional mobility corridors, rendering learned spatial correlations partially outdated even when the global distribution appears stable. The classical remedy of periodic 
retraining is costly, as confirmed in most studies \cite{mahadevan2024cost}, since both temporal and spatial dependencies must be relearned by the underling deployed models \cite{gudepu2024generative, liang2024energy, mahadevan2023cost}, making it impractical for real-time 5G/6G deployments and motivating lightweight drift-mitigation mechanisms.

\subsection{Residual Correction and Feedback-Based Methods}
Post-prediction error correction offers a complementary framework to model retraining. Classical hybrid forecasting established that base-model residuals carry exploitable structure \cite{zhang2003hybrid}, which can be useful for feedback driven mechanisms. ResCAL a residual estimation module presented by the authors in \cite{kim2022rescal} extended this by correcting traffic forecasts in real time using past errors and graph signals \cite{kim2022rescal}, while architecture-agnostic post-processing has 
been revisited as a practical means of improving deployed forecasting models without retraining \cite{liang2026forecast}. Adaptive filtering and Kalman-based methods further formalize this prediction-correction principle for online 
error-driven adjustment \cite{kalman1960new}. More recently, Future-Guided Learning discussed by authors in \cite{gunasekaran2025predictive} demonstrated that dynamic feedback improves time-series forecasting under long-term dependencies and distribution shift \cite{gunasekaran2025predictive}.

Within this family, PID control offers a lightweight and interpretable correction layer. Its proportional term reacts to instantaneous error, the integral term compensates persistent bias, and the derivative term responds to changes in the error trajectory \cite{dormido2021pid, knospe2006pid, 
johnson2005pid}. Additionally, operating entirely at inference time, the PID correction introduces negligible overhead through lightweight computations and a small parameter footprint. Prior works have also shown the potential of this feedback correction for pretrained LSTM-based traffic prediction \cite{john}. While prior studies explored residual correction and feedback-driven forecasting, lightweight PID-based online correction for hierarchical spatiotemporal cellular forecasting under drift remains largely unexplored. In the next section, we present our methodology.

\section{Methodology}
\label{sec:methodology}
This section discusses the PID correction workflow and integration with baseline models.
\subsection{Feedback-Driven Enhancement}

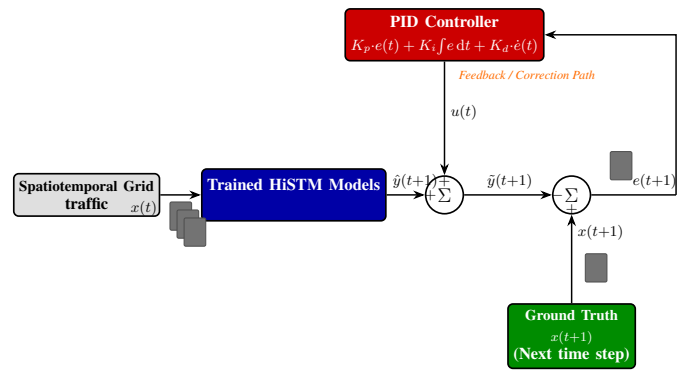
\begin{figure}[htbp]
\centering
\resizebox{\columnwidth}{!}{%
\begin{tikzpicture}[
  font=\fontsize{10}{11}\selectfont\bfseries,
  >=Stealth,
  every path/.style={line width=0.9pt},
  inputblock/.style={
    rectangle, rounded corners=3pt,
    draw=black, line width=1.0pt,
    fill=gray!25,
    minimum width=2.4cm, minimum height=0.9cm,
    align=center,
  },
  modelblock/.style={
    rectangle, rounded corners=3pt,
    draw=black, line width=1.0pt,
    fill=blue!65!black, text=white,
    minimum width=2.6cm, minimum height=1.1cm,
    align=center,
  },
  pidblock/.style={
    rectangle, rounded corners=3pt,
    draw=black, line width=1.0pt,
    fill=red!80!black, text=white,
    minimum width=3.7cm, minimum height=1.1cm,
    align=center,
  },
  gtblock/.style={
    rectangle, rounded corners=3pt,
    draw=black, line width=1.0pt,
    fill=green!55!black, text=white,
    minimum width=2.6cm, minimum height=1.0cm,
    align=center,
  },
  sumjunc/.style={
    circle, draw=black, line width=1.0pt,
    minimum size=0.80cm, fill=white,
  },
  siglabel/.style={
    font=\fontsize{9}{10}\selectfont\bfseries\itshape,
  },
  thumb/.style={
    draw=black!70, fill=black!55,
    minimum width=0.46cm, minimum height=0.60cm,
    rounded corners=1pt, line width=0.5pt,
  },
]

%% ── Nodes ────────────────────────────────────────────────────────

\node[inputblock] (input) at (0.0, 0)
  {\fontsize{9}{10}\selectfont\bfseries Spatiotemporal Grid \\traffic};

\node[modelblock] (model) at (4.4, 0)
  {\textbf{Trained HiSTM Models}\\[2pt]};

\node[sumjunc] (sumcorr) at (7.6, 0) {};
\node[font=\fontsize{13}{13}\selectfont] at (sumcorr) {$\Sigma$};

\node[sumjunc] (sumerr) at (10.3, 0) {};
\node[font=\fontsize{13}{13}\selectfont] at (sumerr) {$\Sigma$};

\node[pidblock] (pid) at (7.6, 3.4)
  {\fontsize{10}{11}\selectfont\bfseries PID Controller\\[3pt]
   \fontsize{9}{10}\selectfont\bfseries
  $K_p{\cdot}e(t)+K_i{\textstyle\int}e\,\mathrm{d}t+K_d{\cdot}\dot{e}(t)$};

\node[gtblock] (gt) at (10.3, -3.0)
  {\fontsize{9}{10}\selectfont\bfseries Ground Truth\\[2pt]
   \fontsize{8}{9}\selectfont\bfseries
   $x(t{+}1)$\\(Next time step)};

%% ── Arrows ───────────────────────────────────────────────────────

%% 1) x(t): Input → Model, enters at lower slot (yshift=-0.22cm)
\draw[->] (input.east)
    -- node[below, siglabel, pos=-0.4]{$x(t)$}
    ([yshift=-0.0cm]model.west);

%% 2) Raw model prediction: Model → correction sum
\draw[->] (model.east)
  -- node[above, siglabel, pos=0.75]{$\hat{y}(t{+}1)$}
    (sumcorr.west);

%% 3) Corrected prediction: first sum → error sum
\draw[->] (sumcorr.east)
    -- node[above, siglabel, pos=0.52]{$\tilde{y}(t{+}1)$}
    (sumerr.west);

%% 4) e(t+1): error sum → right → up → pid.east
\draw (sumerr.east)
    -- ++(1.8,0) coordinate (eR)
       node[above, siglabel, pos=0.75]{$e(t{+}1)$};
\draw[->] (eR) |- (pid.east);

%% 5) u(t): PID output is added to the raw model prediction at the correction sum
\draw[->] (pid.south)
  -- node[right, siglabel, pos=0.45]{$u(t)$}
  (sumcorr.north);

%% 6) x(t+1): GT.north → error sum.south
\draw[->] (gt.north)
    -- node[right, siglabel, pos=0.8]{$x(t{+}1)$}
    (sumerr.south);

%% ── Labels ───────────────────────────────────────────────────────

\node[font=\fontsize{7.5}{8}\selectfont\itshape,
      text=orange!85!red, anchor=north]
  at ([xshift=1.75cm,yshift=-0.05cm]pid.south)
  {Feedback / Correction Path};

\node[font=\fontsize{9}{9}\selectfont\bfseries]
  at ([xshift=-0.30cm]sumcorr.center) {$+$};
\node[font=\fontsize{9}{9}\selectfont\bfseries]
  at ([yshift=0.30cm]sumcorr.center) {$+$};

\node[font=\fontsize{9}{9}\selectfont\bfseries]
  at ([xshift=-0.30cm]sumerr.center) {$-$};
\node[font=\fontsize{9}{9}\selectfont\bfseries]
  at ([yshift=-0.30cm]sumerr.center) {$+$};

%% ── Thumbnails — below the x(t) arrow ───────────────────────────
\node[thumb] at (2.00, -0.45) {};
\node[thumb] at (2.15, -0.62) {};
\node[thumb] at (2.30, -0.79) {};

%% Beside x(t+1) vertical arrow
\node[thumb] at (10.82, -1.55) {};

%% Right of error sum, above e(t+1) line
\node[thumb] at (11.35, 0.62) {};

\end{tikzpicture}}
\caption{PID correction framework integrating  trained HiSTM models with a PID controller to refine traffic predictions.}
\label{fig:pid_framework}
\end{figure}

As previously underscored, our framework builds on pre-trained HiSTM forecasting models and enhances their future time-step predictions through a PID-based output-side correction mechanism. As illustrated in Figure~\ref{fig:pid_framework}, the spatiotemporal input $x(t)$ is first processed by the HiSTM model to produce the future forecast $\hat{y}(t+1)$. This prediction is then combined with the PID corrective signal $u(t)$ to obtain the corrected future prediction $\tilde{y}(t+1)$. The prediction error at the next time step is subsequently computed as the difference between the ground-truth observation $x(t+1)$ and the corrected prediction, yielding $e(t+1)$. This error signal is supplied to the PID controller, whose output is determined by the proportional, integral, and derivative terms \cite{dormido2021pid}, thereby capturing the current error, the accumulated past error, and the variation with respect to the previous error. This forms a closed-loop mechanism where prediction errors continuously refine subsequent traffic forecasts. Additionally, a systematic algorithmic flow is shown in Algorithm~\ref{alg:pid_forecasting} outlining the full PID-enhanced correction workflow, 
taking as input the forecasts produced by the pre-trained HiSTM models and producing corrected predictions. The process begins with an offline Optuna-based hyperparameter search that 
finds the optimal PID gains ($K_p$, $K_i$, $K_d$) by minimizing MAE over 200 trials. 
At each timestep, a correction signal is applied to the raw model prediction to yield 
$\hat{\bar{y}}(t+1) \gets \text{Correct}(\hat{y}(t+1),\ u(t))$. Within the framework, since tuning is done once in offline and reused at inference time, it adds no meaningful computational overhead during deployment.
\begin{algorithm}[htbp]
    \caption{PID-Enhanced HiSTM Forecasting}
    \label{alg:pid_forecasting}
    \KwIn{Pre-trained HiSTM model $f_{\text{HiSTM}}(\cdot)$, clean evaluation data $D_{eval}$, drifted evaluation data $\{D_{\tau}\}$ under various drift types}
    \KwOut{Corrected predictions on clean and drifted data, mitigation metrics per drift type}

    \textit{// Stage 0: Optuna-based PID parameter Tuning} \;
    $K_p, K_i, K_d \gets \text{OptunaTuning}(n\text{\_trials}=200,\ \text{objective}=\text{``minimize MAE''})$ \;

    \textit{// Evaluate on clean data} \;
    Initialize $E_{\text{int}} \gets \mathbf{0}$, $e(0) \gets \mathbf{0}$ \;
    \For{$t \gets 1$ \KwTo $M$}{
        $\hat{y}(t+1) \gets f_{\text{HiSTM}}(x(t))$ \quad \textit{// HiSTM predicts next timestep} \;
        $u(t) \gets K_p \cdot e(t) + K_i \cdot E_{\text{int}} + K_d \cdot \Delta e(t)$ \quad \textit{// PID signal from previous error} \;
        $\hat{\bar{y}}(t+1) \gets \text{Correct}(\hat{y}(t+1),\ u(t))$ \;
        $e(t+1) \gets \hat{\bar{y}}(t+1) - x(t+1)$ \quad \textit{// New error} \;
        $E_{\text{int}} \gets E_{\text{int}} + e(t+1)$ \;
    }
    Compute clean metrics: $\text{MAE\&RMSE}_{\text{HiSTM}}^{clean},\ \text{MAE\&RMSE}_{pid}^{clean}$ \;

    \textit{// Evaluate on drifted data under different drift scenarios} \;
    \For{each drift type $\tau \in$ drift\_types}{
        Reset $E_{\text{int}} \gets \mathbf{0}$, $e(0) \gets \mathbf{0}$;\quad
        Load $D_{\tau} \gets \text{LoadDriftedData}(\tau)$ \;
        \For{$t \gets 1$ \KwTo $M$}{
            $\hat{y}(t+1) \gets f_{\text{HiSTM}}(\tilde{x}(t))$ \quad \textit{// HiSTM predicts next timestep} \;
            $u(t) \gets K_p \cdot e(t) + K_i \cdot E_{\text{int}} + K_d \cdot \Delta e(t)$ \quad \textit{// PID signal from previous error} \;
            $\hat{\bar{y}}(t+1) \gets \text{Correct}(\hat{y}(t+1),\ u(t))$ \;
            $e(t+1) \gets \hat{\bar{y}}(t+1) - \tilde{x}(t+1)$ \quad \textit{// New error} \;
            $E_{\text{int}} \gets E_{\text{int}} + e(t+1)$ \;
        }
        Compute drift metrics: $\text{MAE\&RMSE}_{\text{HiSTM}}^{\tau},\ \text{MAE\&RMSE}_{pid}^{\tau}$ \;
    }

    \KwRet $\hat{\mathbf{y}}^{pid,clean\_data}$, $\{\hat{\mathbf{y}}^{pid,drifted\_data}_{\tau}\}_{\tau}$, metrics
\end{algorithm}
\subsection{PID parameter impact}
The control signal $u(t)$ is generated following Eq.~\ref{eq:pid} below:

\begin{equation}
    u(t) = K_p \cdot e(t) + K_i \cdot \int_0^t e(\tau)\,d\tau + K_d \cdot \frac{de}{dt}
    \label{eq:pid}
\end{equation}

In generating the $u(t)$, the proportional component $K_p$ provides an immediate response to the current error \cite{dormido2021pid}.
This ensures that larger errors result in proportionally larger corrections, enabling rapid reaction to sudden deviations in traffic patterns.
In contrast, the integral component $K_i$ aggregates past errors over time \cite{dormido2021pid}. By incorporating this historical information, it introduces a memory effect that helps eliminate persistent biases caused by sustained drift, thereby improving long-term accuracy.
Finally, the derivative component $K_d$ captures the rate of change of the error \cite{dormido2021pid}.
The trade-offs associated with tuning $K_p$, $K_i$, and $K_d$ are additionally summarized in Table~\ref{Table:pidparameters}, highlighting implications e.g, a high $K_p$ yields fast response but risks oscillation and overshoot, while a high $K_i$ eliminates steady-state error at the cost of potential integral windup. Increasing $K_d$ improves damping but amplifies high-frequency noise.
\renewcommand{\arraystretch}{0.5}
% \begin{table}[t]
% \centering
% \caption{Effects of PID Gains ($K_p$, $K_i$, $K_d$)}
% \label{Table:pidparameters}
% \resizebox{\columnwidth}{!}{
% \begin{tabular}{p{2cm} p{2cm} p{4.5cm} p{4.5cm}}
% \hline
% \textbf{Component} & \textbf{Gain Level} & \textbf{Advantages} & \textbf{Disadvantages} \\
% \hline

% \multirow{2}{*}{$K_p$} 
% & High 
% & Fast response; reduces steady-state error; suitable for sudden traffic changes \cite{Astrom2005, Johnson2005} 
% & May cause oscillations and overshooting; amplifies measurement noise \cite{Bai2018, Astrom2005} \\

% & Low 
% & Stable, smooth adjustments; reduces oscillations 
% & Slow response; larger steady-state error \cite{Johnson2005} \\

% \hline

% \multirow{2}{*}{$K_i$} 
% & High 
% & Eliminates steady-state error; corrects persistent bias; effective for long-term drift \cite{Bingi2020, Johnson2005} 
% & Can cause oscillations and instability; risk of integral windup \\

% & Low 
% & Prevents windup; improves stability 
% & Persistent steady-state error; slower error correction \cite{Johnson2005} \\

% \hline

% \multirow{2}{*}{$K_d$} 
% & High 
% & Improves damping; reduces overshoot; enhances transient response \cite{Rojas2021, Astrom2005} 
% & Amplifies high-frequency noise; may cause sluggish response to real signals \cite{Alfaro2016, Tan2012} \\

% & Low 
% & Less sensitivity to noise; smoother control in noisy environments \cite{Tan2012} 
% & Reduced damping; oscillations and overshoot may persist \cite{Rojas2021} \\

% \hline
% \end{tabular}}
% \end{table}

\begin{table}[t]
    \centering
    \caption{Effects of PID Gains on Controller operation \cite{johnson2005pid,bai2019classical,bingi2020fractional,alfaro2016model }}
    \label{Table:pidparameters}
    \setlength{\tabcolsep}{3pt}
    \renewcommand{\arraystretch}{0.9}
    \begin{tabularx}{\columnwidth}{
      >{\bfseries\centering\arraybackslash}p{0.8cm}
      >{\centering\arraybackslash}p{0.8cm}
      >{\RaggedRight\arraybackslash}X
      >{\RaggedRight\arraybackslash}X
    }
    \toprule
    \textbf{Term} & \textbf{Level} & \textbf{Merit} & \textbf{Drawback} \\
    \midrule

    \multirow{2}{*}{$K_p$}
      & \textit{High} & Fast response; reduces steady-state error
                      & Oscillations and overshoot \\
      & \textit{Low}  & Stable, smooth adjustments
                      & Slow response; larger steady-state error \\
    \midrule

    \multirow{2}{*}{$K_i$}
      & \textit{High} & Eliminates steady-state error; corrects persistent bias
                      & Oscillations; risk of integral windup \\
      & \textit{Low}  & Avoids integrator-driven instability
                      & Persistent steady-state error \\
    \midrule

    \multirow{2}{*}{$K_d$}
      & \textit{High} & Improves damping; reduces overshoot
                      & Amplifies high-frequency noise \\
      & \textit{Low}  & Less noise sensitivity; smoother control
                      & Reduced damping; overshoot may persist \\
    \bottomrule
    \end{tabularx}
\end{table}
\renewcommand{\arraystretch}{1.0}
\subsection{PID per cell}
Rather than applying a single global PID setting across all cells, our framework performs per-cell PID parameter tuning, accounting for the fact that each cell serves a varying number of users and experiences distinct traffic and interference patterns. Moreover, this per-cell approach is well-supported in literature works such as in \cite{Yinghong_WEN}, where they employ cell-specific association strategies in heterogeneous networks (HetNets) to ensure equitable user service, while related 5G research in \cite{Shami} takes this further by assigning each small cell a tailored biasing value to balance load and improve throughput, both of which align closely with our per-cell PID parameter tuning approach.

For each independent cell $c$, Optuna a hyperparameter optimization framework \cite{akiba2019optuna} was applied to identify the optimal PID gain 
parameters $\theta = (K_p, K_i, K_d)$ within a budget of $T = 200$ optuna trials. A dedicated 
Optuna study was instantiated per cell, ensuring that the Tree-structured Parzen Estimator (TPE) \cite{bergstra2011algorithms} surrogate model and pruning 
state remained isolated across cells.

\begin{equation}
    \mathcal{L}^{(c)}(\theta_t) = \frac{1}{N_c} \sum_{\tau=1}^{N_c} 
    \left| y_{\text{true}}^{(c)}(\tau) - y_{\text{pred}}^{(c)}(\tau;\,\theta_t) \right|
    \label{eq:optuna}
\end{equation}
$N_c = \min\left(|\mathcal{T}_c|, 200\right)$
The per-cell optimization problem was formulated as:
\begin{equation}
    \theta_c^* = \underset{\theta \in \Theta}{\arg\min}\; \mathcal{L}^{(c)}(\theta)
\end{equation}
where $\mathcal{L}^{(c)}(\theta)$ denotes the Mean Absolute Error (MAE) see Eq.~(\ref{eq:optuna}).
To optimize this, optuna framework employed the Tree-structured Parzen Estimator, which 
iteratively refined the sampling strategy based on the history of prior evaluations 
$\left\{(\theta_j,\, \mathcal{L}^{(c)}(\theta_j))\right\}_{j=1}^{t-1}$. Specifically, 
TPE fitted two separate density models over the observed configurations: $\ell(\theta)$ 
over well-performing trials and $g(\theta)$ over poor ones. Each subsequent candidate 
was then drawn according to:

\begin{equation}
    \theta_{t+1} \sim \frac{\ell(\theta)}{g(\theta)}
\end{equation}

thereby optimizing the search toward regions of $\Theta$ associated with lower MAE. 
To further improve the effective use of the evaluation budget, a Median Pruner was 
activated after five startup trials, allowing unpromising configurations to be 
terminated before completing their full evaluation. Upon exhausting all $T = 200$ 
trials, the best-performing parameter set $\theta_c^* = (K_p^*,\, K_i^*,\, K_d^*)$ 
was retained for each cell $c$, replacing any previously stored result only when
 $\mathcal{L}^{(c)}(\theta_c^*) < \mathcal{L}^{(c)}_{\text{prev}}$.

\renewcommand{\arraystretch}{0.5}
\begin{table}[t]
\centering
\scriptsize
\setlength{\tabcolsep}{4pt}
\caption{Summary of the key model parameters and experimental setup.}
\label{tab:histm_summary}
\begin{tabular*}{\columnwidth}{@{\extracolsep{\fill}}lcc}
\toprule
Model & Parameters & Size (MB) \\
\midrule
HiSTM        & 33,794 & 0.139 \\
HiSTM\_Nested & 39,139 & 0.164 \\
\midrule
Architecture     & \multicolumn{2}{c}{2 CNN+Mamba encoders, 32 channels; Nested adds memory} \\
Input Shape      & \multicolumn{2}{c}{$L{=}6$, patch $K{=}11{\times}11$, stride 5} \\
Prediction Horizon & \multicolumn{2}{c}{1-step ahead (10\,min); up to 6-step autoregressive} \\
Data Split       & \multicolumn{2}{c}{Train / Val / Test: 70\% / 15\% / 15\%} \\
Preprocessing    & \multicolumn{2}{c}{Min-Max scaling to $[0,1]$} \\
\midrule
GPU              & \multicolumn{2}{c}{NVIDIA A100 80\,GB} \\
CPU / RAM        & \multicolumn{2}{c}{64 cores / 512\,GB} \\
CUDA             & \multicolumn{2}{c}{12.4} \\
Batch Size       & \multicolumn{2}{c}{\textit{128} (eval: 512)} \\
Optimizer        & \multicolumn{2}{c}{Adam ($\text{lr}{=}10^{-4}$)} \\
\bottomrule
\end{tabular*}
\end{table}
\renewcommand{\arraystretch}{1.0}

\subsection{Computational Complexity Analysis} 
The temporal complexity per inference step is dominated by the based model forward pass: $O(L \times H^2)$, where $L$ is the number of layers and $H$ is the hidden dimension. The PID correction overhead is strictly $O(c_{out})$, corresponding to three simple scalar operations per output cell (proportional gain: $K_p \cdot e_t$; integral accumulation: $K_i \cdot E_{int}$; derivative term: $K_d \cdot \Delta e_t$). The spatial complexity is minimal as only 3 parameters per output cell ($K_p, K_i, K_d$) must be stored, amounting to $O(3 \times c_{out})$ memory, which is negligible relative to the pre-trained neural network weights. Critically, the PID framework requires \textit{no backpropagation} during inference, the model weights remain frozen and only the correction signal is computed feedforward, further minimizing computational burden. 
\renewcommand{\arraystretch}{0.5}

\begin{table*}[t]
\centering
\caption{Optimised PID parameters and MAE per cell}
\label{tab:pid_params}
\setlength{\tabcolsep}{5pt}
\renewcommand{\arraystretch}{1.05}
\begin{tabular}{
  c
  S[table-format=1.6]
  S[table-format=1.6]
  S[table-format=1.6]
  S[table-format=2.4]
  S[table-format=1.6]
  S[table-format=1.6]
  S[table-format=1.6]
  S[table-format=2.4]
  c
}
\toprule
& \multicolumn{4}{c}{\textbf{histm}} & \multicolumn{4}{c}{\textbf{histm\_nested}} & \\
\cmidrule(lr){2-5}\cmidrule(lr){6-9}
\textbf{Cell} & {$K_p$} & {$K_i$} & {$K_d$} & {MAE} & {$K_p$} & {$K_i$} & {$K_d$} & {MAE} & \textbf{Trials} \\
\midrule
155 & 0.000124 & 0.000129 & 0.018972 & 12.6549 & 0.006588 & 0.000824 & 0.110250 & 13.5583 & 200 \\
160 & 0.026881 & 0.000164 & 0.000143 & 12.8134 & 0.027407 & 0.001473 & 0.450456 & 16.7808 & 200 \\
\textbf{165} & \textbf{0.048205} & \textbf{0.001748} & \textbf{0.012799} & \textbf{6.8594} & \textbf{0.011059} & \textbf{0.001885} & \textbf{0.506242} & \textbf{9.0165} & \textbf{200} \\
170 & 0.000167 & 0.000027 & 0.040233 & 15.7955 & 0.003134 & 0.000722 & 0.112583 & 17.3439 & 200 \\
\midrule
305 & 0.000412 & 0.000406 & 0.003810 & 10.8859 & 0.001044 & 0.001641 & 0.030686 & 11.4189 & 200 \\
310 & 0.000127 & 0.000176 & 0.081113 & 15.5547 & 0.000119 & 0.002214 & 0.001423 & 17.1430 & 200 \\
315 & 0.071574 & 0.000129 & 0.041672 & 19.2837 & 0.000812 & 0.001763 & 0.019767 & 20.4128 & 200 \\
320 & 0.000283 & 0.000104 & 0.008895 & 23.3264 & 0.002849 & 0.000125 & 0.000467 & 24.4297 & 200 \\
\midrule
455 & 0.000130 & 0.000454 & 0.001213 & 16.1323 & 0.010893 & 0.003172 & 0.000961 & 16.7138 & 200 \\
460 & 0.000403 & 0.000097 & 0.055795 & 14.6235 & 0.000485 & 0.001880 & 0.098531 & 15.5530 & 200 \\
465 & 0.000268 & 0.000386 & 0.009343 & 19.4919 & 0.019177 & 0.001792 & 0.035751 & 20.3202 & 200 \\
470 & 0.001822 & 0.000692 & 0.001288 & 18.6255 & 0.002462 & 0.000599 & 0.090479 & 18.7821 & 200 \\
\midrule
\textbf{605} & \textbf{0.014860} & \textbf{0.000090} & \textbf{0.000339} & \textbf{6.6326} & \textbf{0.001683} & \textbf{0.000158} & \textbf{0.081255} & \textbf{6.8698} & \textbf{200} \\
\textbf{610} & \textbf{0.001117} & \textbf{0.001139} & \textbf{0.001745} & \textbf{5.0714} & \textbf{0.007529} & \textbf{0.001866} & \textbf{0.076038} & \textbf{5.8360} & \textbf{200} \\
615 & 0.002054 & 0.000087 & 0.033782 & 17.9314 & 0.002123 & 0.001476 & 0.127340 & 18.4561 & 200 \\
620 & 0.000238 & 0.000975 & 0.000127 &  7.3313 & 0.000152 & 0.001627 & 0.049276 &  8.0868 & 200 \\
\bottomrule
\end{tabular}
\end{table*}
\renewcommand{\arraystretch}{1.0}
\renewcommand{\arraystretch}{0.5}

\begin{table*}[t]
\centering
\caption{Drift scenarios and their impact}
\label{table:selected_drifts}
\setlength{\tabcolsep}{4pt}
\renewcommand{\arraystretch}{0.9}
\small
\begin{tabular*}{\linewidth}{
  @{}
  >{\centering\arraybackslash}p{0.35cm}
  >{\centering\arraybackslash}p{1.6cm}
  >{\centering\arraybackslash}p{1.4cm}
  >{\centering\arraybackslash}p{2.2cm}
  p{\dimexpr\linewidth-0.35cm-1.6cm-1.4cm-2.2cm-10\tabcolsep-2\arrayrulewidth\relax}
  @{}
}
\toprule
\textbf{\#} & \textbf{Category} & \textbf{Type} & \textbf{Scope} & \textbf{Mechanism \& Impact} \\
\midrule

1 & Hotspot & Linear \cite{Gama} Local &
  Regional Block; gradual &
  A high-traffic zone slowly grows in a random region, exceeding historical levels. Models new cell tower activation or gradual load-balancing. \\[4pt]

2 & Hotspot & Sudden \cite{Gama} Global &
  Global; instantaneous &
  Network-wide traffic surge across all cells simultaneously. Emulates flash-crowd events or competitor failure redirecting subscribers. \\[4pt]

3 & Joint ST & Sudden \cite{Gama} &
  Global (rotating); instantaneous &
  A daily-rotating hotspot appears instantly, introducing coupled spatial-temporal non-stationarity. Models abrupt urban mobility shifts. \\[4pt]

4 & Joint ST & Linear &
  Global (rotating); gradual &
  The rotating hotspot grows linearly from zero to full intensity. Models slow urban-mobility transformation, e.g., new transit routing. \\[4pt]

5 & Joint ST & Recurring \cite{Gama} &
  Global (rotating); sinusoidal &
  The rotating hotspot pulses sinusoidally; location and intensity oscillate simultaneously, requiring adaptive modeling. \\

\bottomrule
\end{tabular*}
\end{table*}
\renewcommand{\arraystretch}{1.0}

\section{Experimental setup}
\label{sec:experimental}
In this section, we provide an overview of the trained model configurations,  datasets \&\ preprocessing, drift injection, and evaluation mechanisms applied to measure performance.
\begin{figure}
\centerline{\includegraphics[width=\columnwidth]{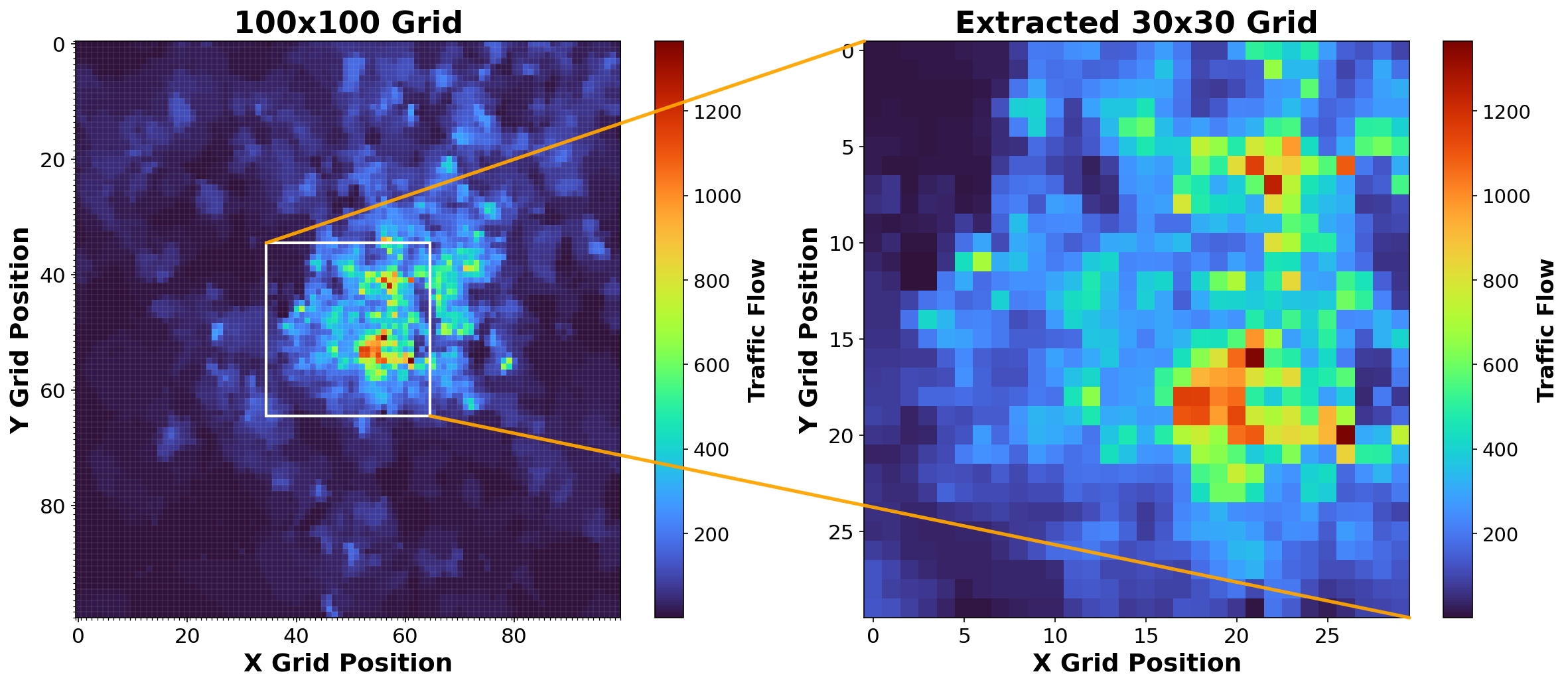}}
    \caption{Spatiotemporal grid traffic visualization showing full 100×100 full grid (left) and an extracted 30×30 subgrid (right)}
    \label{fig:grid}
\end{figure}
\subsection{Model configuration}
Table~\ref{tab:histm_summary} highlights key model parameters for the two trained models.
HiSTM used 33,794 parameters and HiSTM\_Nested extends this
with a memory decay mechanism~\cite{nested}, totaling 39,139 parameters.
Both models operated on $11{\times}11$ spatial patches with a
stride of 5, across CNN+Mamba
encoder layers. All experiments were conducted on an
NVIDIA A100 80\,GB GPU under CUDA 12.4, using the Adam
optimizer and early stopping to retain the best checkpoint.

\subsection{Dataset and Preprocessing}
An open source 5G spatiotemporal traffic dataset provided in
\cite{barlacchi2015multi} was applied in our implementation, providing a comprehensive view of urban activities in
Milan city \cite{hussien2025machine}, spanning a structured
urban grid as illustrated by the $100 \times 100$ grid
(on the left of Figure~\ref{fig:grid}), on the right showing
the extracted region applied during the evaluation phase.
During model training, the dataset was chronologically partitioned into training (70\%),validation (15\%), and test (15\%) sets to preserve temporal order and prevent data leakage. Input features and target values normalized to $[0,1]$ with Min-Max scaling as underscored in Table~\ref{tab:histm_summary}.

\subsection{Evaluation}

After saving the best model checkpoints, evaluation was conducted in pipeline stages to provide a comprehensive assessment of our proposed feedback framework. The dataset applied in evaluation was a $30 \times 30$ region of interest earlier on extracted form the main grid as (depicted in Figure~\ref{fig:grid}). 
Initially, pre-trained models were evaluated on the extracted dataset to establish baseline performance, followed by incorporation of the PID controllers to enhance prediction performance, enabling a comparative analysis between the standalone baseline models 
and the enhancement with the \textbf{control feedback mechanisms}. Algorithm~\ref{alg:pid_forecasting} earlier presented further provides a guided step throughout the evaluation.
The evaluation comprised three stages:

\begin{itemize}
    \item \textbf{Optuna-based hyperparameter tuning:} Optimal PID parameters 
    ($K_p, K_i, K_d$) were identified for each cell. Due to page limit requirements, we provide a representative 
    subset of 16 cells shown in Table~\ref{tab:pid_params}, reporting the optimized 
    parameters $(K_p^*, K_i^*, K_d^*)$ and the corresponding \textbf{MAE} achieved per cell 
    for both \textit{HiSTM} and \textit{HiSTM\_Nested} models, each evaluated over 
    $T = 200$ optuna trials. Each trial $t \in \{1, \dots, T\}$ sampled a candidate configuration $\theta_t = (K_p, K_i, K_d)$ from the bounded wide parameter search space, where $K_p \sim \log\mathcal{U}(10^{-4}, 10.0)$, $K_i \sim \mathcal{U}(0, 1.0)$, and $K_d \sim \mathcal{U}(0, 2.0)$. Notably, cells such as 610 and 605 emerge as 
    top performers, mainly attributed to their steady traffic patterns in less busy areas, 
    whereas cells such as  320 residing in high traffic-volatility zones 
    (e.g., central business districts) registered comparatively higher errors.

    \item \textbf{Baseline evaluation on clean data:} Models were assessed on clean data 
    to establish baseline PID effectiveness.

    \item \textbf{Evaluation under data drift scenarios:} Robustness was assessed across 
    different drift scenarios, where the \textbf{mitigation percentage} at each 
    scale quantified how effectively the PID controller mitigates drift-induced errors.
\end{itemize}

During drift evaluation, we injected drift scenarios into the traffic data to emulate realistic changes commonly observed in network traffic environments. Formally, the drifted data emulated as
\begin{equation}
X_{t}^{\text{drift}} = X_t + \delta_t
\end{equation}
where $X_t$ denotes the original spatiotemporal traffic observation at time $t$, and $\delta_t$ represents the injected perturbation that introduces pattern deviations. In the workflow, drift intensity was controlled through severity (\emph{low}, \emph{mid}, \emph{high}) associated with nominal gain parameters of $0.05$, $0.15$, and $0.30$, respectively, consistent with prior works that treat concept drift as a quantitative change in both the speed and amount of change between concepts \cite{moulton2018clustering}. The realized drift magnitude was quantified from the generated drifted tensors using the mean absolute deviation (MAD), root mean squared deviation (RMSD), relative drift strength, and the fraction of changed samples. The drift scenarios considered are summarized in Table~\ref{table:selected_drifts}, together with their practical implications on network traffic. This additional work-flow enabled a systematic assessment of how both the standalone models and the PID-enhanced framework respond to such perturbations. Additionally, by incorporating drift changes to reflect patterns such as user mobility, and emerging traffic patterns \cite{tziouvaras2025towards}, our framework reflects a more realistic approach and paves the way for a comprehensive evaluation of the framework's robustness under evolving data dynamics \cite{singh2013quantifying}, critical in future 6G.

% -------------------------------------------------------
%  Required packages in preamble:
%    \usepackage{tikz}
%    \usepackage{xcolor}
%    \usepackage{pgfplots}
%    \pgfplotsset{compat=1.18}
%
%  Place these \definecolor lines in your preamble:
%    \definecolor{colBL}    {RGB}{ 70,130,180}
%    \definecolor{colBLPID} {RGB}{ 30, 70,140}
%    \definecolor{colBLD}   {RGB}{210, 80, 45}
%    \definecolor{colBLDPID}{RGB}{220,170,  0}
% -------------------------------------------------------

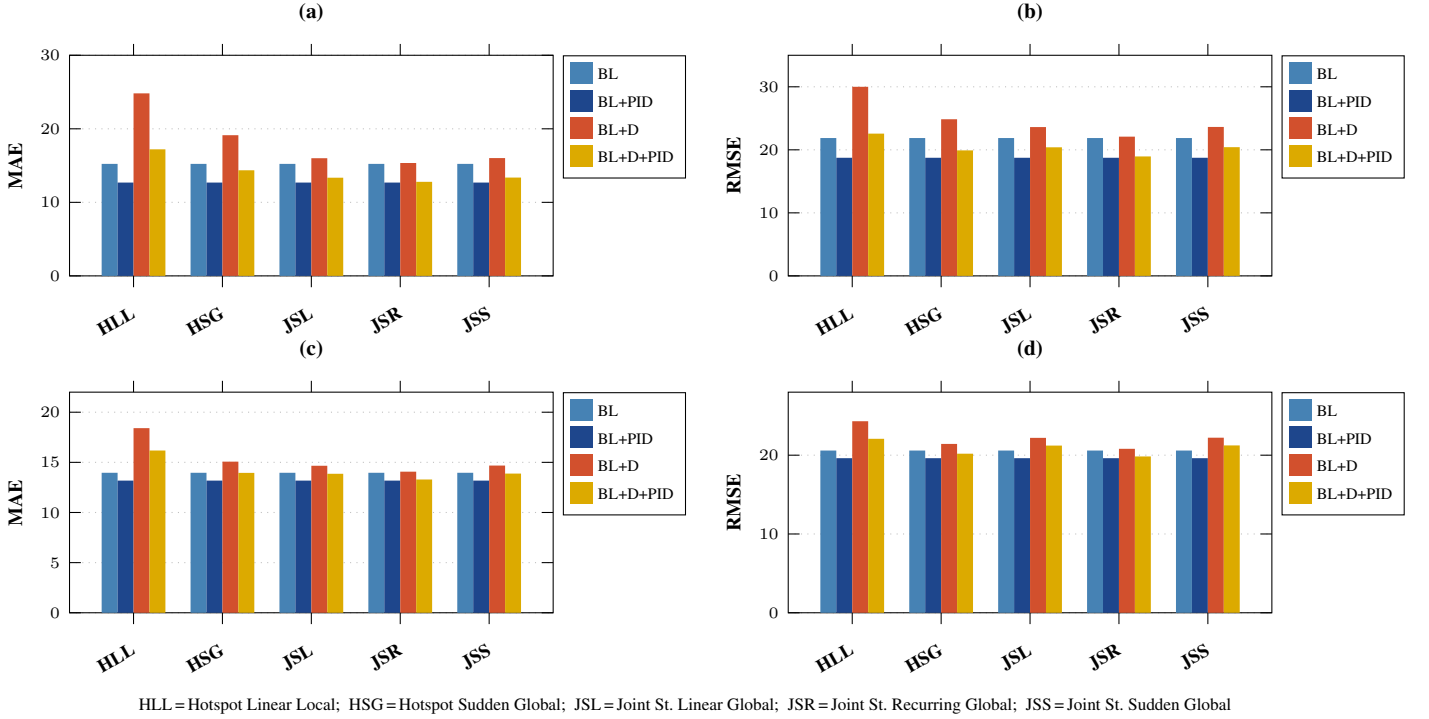
\begin{figure*}[htbp]
\centering

\pgfplotsset{
  barplot/.style={
    ybar,
    bar width        = 6pt,
    width            = 0.44\textwidth,
    height           = 4.50cm,
    enlarge x limits = 0.18,
    ymajorgrids      = true,
    grid style       = {dotted, gray!50},
    axis line style  = {thin, black},
    tick style       = {thin, black},
    xtick            = data,
    % CHANGED: abbreviated labels, larger font, less rotation
    xticklabel style = {
      font     = \fontsize{7}{7.5}\selectfont\bfseries,
      align    = right,
      rotate   = 30,
      anchor   = north east,
    },
    % CHANGED: reduce bottom margin so x labels don't eat vertical space
    xtick align      = outside,
    xticklabel shift = 2pt,
    yticklabel style  = {font=\fontsize{6.5}{6.5}\selectfont},
    legend image code/.code={%
      \draw[draw=none, fill=#1]
        (0cm,-0.10cm) rectangle (0.30cm,0.18cm);
    },
    legend style = {
      font          = \fontsize{6}{6.5}\selectfont,
      at            = {(1.02,1.00)},
      anchor        = north west,
      draw          = black,
      line width    = 0.4pt,
      inner sep     = 2.5pt,
      row sep       = 0.5pt,
    },
    legend cell align = left,
    title style = {
      font   = \fontsize{8}{9}\selectfont\bfseries,
      yshift = 3pt,
    },
    ylabel style = {font=\fontsize{7.5}{7.5}\selectfont\bfseries},
    every axis/.append style = {line width=0.5pt},
  }
}%
%
% CHANGED: \vspace reduced between rows; tabular row sep tightened
%
\begin{tabular}{@{}cc@{}}
%
% ── Row 1 ─────────────────────────────────────────────────────────────
%
% (a) MAE – HiSTM-Nested
\begin{tikzpicture}
\begin{axis}[
  barplot,
  title   = {(a)},
  ylabel  = {MAE},
  ymin    = 0, ymax = 30,
  symbolic x coords = {HLL, HSG, JSL, JSR, JSS},
  xticklabels = {HLL, HSG, JSL, JSR, JSS},
  legend entries = {BL, BL+PID, BL+D, BL+D+PID},
]
\addplot[fill=colBL,     draw=none, bar shift=-9pt] coordinates
  {(HLL,15.23)(HSG,15.23)(JSL,15.23)(JSR,15.23)(JSS,15.23)};
\addplot[fill=colBLPID,  draw=none, bar shift=-3pt] coordinates
  {(HLL,12.69)(HSG,12.69)(JSL,12.69)(JSR,12.69)(JSS,12.69)};
\addplot[fill=colBLD,    draw=none, bar shift= 3pt] coordinates
  {(HLL,24.82)(HSG,19.13)(JSL,16.00)(JSR,15.35)(JSS,16.02)};
\addplot[fill=colBLDPID, draw=none, bar shift= 9pt] coordinates
  {(HLL,17.21)(HSG,14.36)(JSL,13.35)(JSR,12.79)(JSS,13.37)};
\end{axis}
\end{tikzpicture}
&
% (b) RMSE – HiSTM-Nested
\begin{tikzpicture}
\begin{axis}[
  barplot,
  title   = {(b)},
  ylabel  = {RMSE},
  ymin    = 0, ymax = 35,
  symbolic x coords = {HLL, HSG, JSL, JSR, JSS},
  xticklabels = {HLL, HSG, JSL, JSR, JSS},
  legend entries = {BL, BL+PID, BL+D, BL+D+PID},
]
\addplot[fill=colBL,     draw=none, bar shift=-9pt] coordinates
  {(HLL,21.87)(HSG,21.87)(JSL,21.87)(JSR,21.87)(JSS,21.87)};
\addplot[fill=colBLPID,  draw=none, bar shift=-3pt] coordinates
  {(HLL,18.74)(HSG,18.74)(JSL,18.74)(JSR,18.74)(JSS,18.74)};
\addplot[fill=colBLD,    draw=none, bar shift= 3pt] coordinates
  {(HLL,30.00)(HSG,24.85)(JSL,23.61)(JSR,22.09)(JSS,23.63)};
\addplot[fill=colBLDPID, draw=none, bar shift= 9pt] coordinates
  {(HLL,22.57)(HSG,19.91)(JSL,20.40)(JSR,18.95)(JSS,20.42)};
\end{axis}
\end{tikzpicture}
\\[-1.0em] 
%
% ── Row 2 ─────────────────────────────────────────────────────────────
%
% (c) MAE – HiSTM
\begin{tikzpicture}
\begin{axis}[
  barplot,
  title   = {(c)},
  ylabel  = {MAE},
  ymin    = 0, ymax = 22,
  symbolic x coords = {HLL, HSG, JSL, JSR, JSS},
  xticklabels = {HLL, HSG, JSL, JSR, JSS},
  legend entries = {BL, BL+PID, BL+D, BL+D+PID},
]
\addplot[fill=colBL,     draw=none, bar shift=-9pt] coordinates
  {(HLL,13.96)(HSG,13.96)(JSL,13.96)(JSR,13.96)(JSS,13.96)};
\addplot[fill=colBLPID,  draw=none, bar shift=-3pt] coordinates
  {(HLL,13.18)(HSG,13.18)(JSL,13.18)(JSR,13.18)(JSS,13.18)};
\addplot[fill=colBLD,    draw=none, bar shift= 3pt] coordinates
  {(HLL,18.41)(HSG,15.07)(JSL,14.66)(JSR,14.07)(JSS,14.68)};
\addplot[fill=colBLDPID, draw=none, bar shift= 9pt] coordinates
  {(HLL,16.18)(HSG,13.95)(JSL,13.86)(JSR,13.29)(JSS,13.88)};
\end{axis}
\end{tikzpicture}
&
% (d) RMSE – HiSTM
\begin{tikzpicture}
\begin{axis}[
  barplot,
  title   = {(d)},
  ylabel  = {RMSE},
  ymin    = 0, ymax = 28,
  symbolic x coords = {HLL, HSG, JSL, JSR, JSS},
  xticklabels = {HLL, HSG, JSL, JSR, JSS},
  legend entries = {BL, BL+PID, BL+D, BL+D+PID},
]
\addplot[fill=colBL,     draw=none, bar shift=-9pt] coordinates
  {(HLL,20.59)(HSG,20.59)(JSL,20.59)(JSR,20.59)(JSS,20.59)};
\addplot[fill=colBLPID,  draw=none, bar shift=-3pt] coordinates
  {(HLL,19.62)(HSG,19.62)(JSL,19.62)(JSR,19.62)(JSS,19.62)};
\addplot[fill=colBLD,    draw=none, bar shift= 3pt] coordinates
  {(HLL,24.31)(HSG,21.43)(JSL,22.20)(JSR,20.81)(JSS,22.22)};
\addplot[fill=colBLDPID, draw=none, bar shift= 9pt] coordinates
  {(HLL,22.08)(HSG,20.19)(JSL,21.22)(JSR,19.84)(JSS,21.24)};
\end{axis}
\end{tikzpicture}
\end{tabular}

\vspace{0.1em}
% Abbreviation legend line
{\fontsize{6.5}{7.5}\selectfont
HLL\,=\,Hotspot Linear Local;\enspace
HSG\,=\,Hotspot Sudden Global;\enspace
JSL\,=\,Joint St.\ Linear Global;\enspace
JSR\,=\,Joint St.\ Recurring Global;\enspace
JSS\,=\,Joint St.\ Sudden Global}

\caption{MAE and RMSE performance of HiSTM and HiSTM\_Nested under baseline, drift, and PID-corrected conditions across five drift scenarios. (a) HiSTM\_Nested MAE. (b) HiSTM\_Nested RMSE. (c) HiSTM MAE. (d) HiSTM RMSE.}
\label{fig:combined_metrics}
\end{figure*}
\subsection{Evaluation metrics}

Within the architectural work-flow, performance was evaluated using complementary metrics adopted in traffic prediction, Mean Absolute Error (MAE) and Root Mean Squared Error (RMSE) \cite{ chai2014root}.
MAE measuring the average magnitude of prediction errors without considering their direction, where $y_i$ and $\hat{y}i$ denote the actual and predicted values, respectively (See Eq.~(\ref{eq:metrics}) below):

\begin{equation}
\mathrm{MAE} = \frac{1}{n}\sum_{i=1}^{n}|y_i-\hat{y}_i|
\qquad
\mathrm{RMSE} = \sqrt{\frac{1}{n}\sum_{i=1}^{n}(y_i-\hat{y}_i)^2}
\label{eq:metrics}
\end{equation}
Its equal treatment of all deviations makes it an interpretable measure of typical prediction accuracy in the original units of the data.
RMSE defined as in, Eq.~(\ref{eq:metrics}) penalizes larger errors more heavily due to the squared term, making it particularly sensitive to peak deviations arising from sudden congestion or demand spikes in network traffic \cite{chai2014root}.
Together, these metrics provided complementary perspectives on model performance, enabling a robust and holistic assessment of performance. In the next section, we present an overview of our results obtained.

\section{Results Discussion}
\label{sec:analysis}
Results are discussed below, quantitatively showing the effectiveness of the PID correction framework in reducing prediction errors across both model architectures and drift scenarios.

Figure~\ref{fig:combined_metrics} presents the average performance across cells, providing a comparative analysis of MAE and RMSE under all drift scenarios for both \textit{HiSTM\_Nested} and \textit{HiSTM} models across four experimental conditions: Baseline (BL), Baseline with PID correction (BL+PID), Baseline under drift (BL+D), and Baseline under drift with PID correction (BL+D+PID). The results clearly demonstrate the detrimental impact of traffic drift on prediction performance. This effect is evident from the BL+D condition (red bars), which consistently exhibits the highest error values across nearly all scenarios and architectures. Results further show that integrating the PID correction mechanism effectively mitigates this degradation, as the BL+D+PID condition (orange bars) consistently reduces both MAE and RMSE relative to the (BL+D) baselines. This trend remains consistent across both model architectures, indicating that the PID framework generalizes well across different models.

Additionally, Figure~\ref{fig:pid_boxplots_four_panel} presents a fine-grained box plot analysis to further examine the distribution of prediction errors across the different drift scenarios for both \textit{HiSTM} and \textit{HiSTM\_Nested} models. The box plots provide additional insight into the central tendency, variability, spread, and consistency across individual observations. A general shift toward lower error distributions is observed when PID correction is applied under drifted conditions BL+D+PID compared to BL+D, as reflected by the tendency toward lower median values and more compact distributions across both MAE and RMSE metrics. This trend is observed across most evaluated scenarios, depicting that the PID framework not only reduces prediction errors but also improves stability by reducing variability across observations.
% ============================================================
%  Improved four-panel boxplot figure
%  Drop-in replacement for a two-column LaTeX document.
%  Preamble requirements:
%    \usepackage{pgfplots}
%    \usepgfplotslibrary{groupplots,statistics}
%    \pgfplotsset{compat=1.18}
%    \usepackage{xcolor}
%    \usepackage{tikz}
% ============================================================

% ── Colour palette ──────────────────────────────────────────
\definecolor{pidblue}{RGB}{18,59,109}        % dark navy – raw drift
\definecolor{pidorange}{RGB}{245,166,35}     % amber   – PID-mitigated
\definecolor{gridgray}{RGB}{210,210,210}     % soft grid lines
\definecolor{axgray}{RGB}{80,80,80}          % axis / tick label colour

% ── Shared box-plot data macros ─────────────────────────────
%  Each \addplot pair:  left  = pidblue (raw),  right = pidorange (PID)
%  Positions: group centres at 1, 4, 7, 10, 13
%             offset ±0.32 so boxes never touch

\newcommand{\BoxPairHistmMae}{%
  % HLL
  \addplot+[boxplot prepared={draw position=0.68,
      lower whisker=11.124945, lower quartile=15.559076,
      median=19.163791, upper quartile=21.307686, upper whisker=25.386680},
    fill=pidblue, fill opacity=0.80, draw=pidblue!60!black, line width=0.7pt] coordinates {};
  \addplot+[boxplot prepared={draw position=1.32,
      lower whisker=8.306700,  lower quartile=12.143831,
      median=16.796364, upper quartile=18.979340, upper whisker=25.335239},
    fill=pidorange, fill opacity=0.80, draw=pidorange!60!black, line width=0.7pt] coordinates {};
  % HSG
  \addplot+[boxplot prepared={draw position=3.68,
      lower whisker=6.382489, lower quartile=11.474146,
      median=16.342447, upper quartile=18.539089, upper whisker=23.914993},
    fill=pidblue, fill opacity=0.80, draw=pidblue!60!black, line width=0.7pt] coordinates {};
  \addplot+[boxplot prepared={draw position=4.32,
      lower whisker=4.878308,  lower quartile=9.718416,
      median=13.939519, upper quartile=18.143718, upper whisker=23.865347},
    fill=pidorange, fill opacity=0.80, draw=pidorange!60!black, line width=0.7pt] coordinates {};
  % JSS
  \addplot+[boxplot prepared={draw position=6.68,
      lower whisker=5.371039, lower quartile=10.401469,
      median=15.252904, upper quartile=19.502078, upper whisker=26.721301},
    fill=pidblue, fill opacity=0.80, draw=pidblue!60!black, line width=0.7pt] coordinates {};
  \addplot+[boxplot prepared={draw position=7.32,
      lower whisker=4.372462,  lower quartile=9.030285,
      median=13.075801, upper quartile=18.904555, upper whisker=26.668924},
    fill=pidorange, fill opacity=0.80, draw=pidorange!60!black, line width=0.7pt] coordinates {};
  % JSL
  \addplot+[boxplot prepared={draw position=9.68,
      lower whisker=5.352845, lower quartile=10.373360,
      median=15.251040, upper quartile=19.417662, upper whisker=26.721301},
    fill=pidblue, fill opacity=0.80, draw=pidblue!60!black, line width=0.7pt] coordinates {};
  \addplot+[boxplot prepared={draw position=10.32,
      lower whisker=4.357375,  lower quartile=9.030285,
      median=13.075801, upper quartile=18.901081, upper whisker=26.668924},
    fill=pidorange, fill opacity=0.80, draw=pidorange!60!black, line width=0.7pt] coordinates {};
  % JSR
  \addplot+[boxplot prepared={draw position=12.68,
      lower whisker=5.157447, lower quartile=10.048878,
      median=15.143215, upper quartile=18.583813, upper whisker=23.891506},
    fill=pidblue, fill opacity=0.80, draw=pidblue!60!black, line width=0.7pt] coordinates {};
  \addplot+[boxplot prepared={draw position=13.32,
      lower whisker=4.184772,  lower quartile=8.771107,
      median=12.603029, upper quartile=18.309437, upper whisker=23.840756},
    fill=pidorange, fill opacity=0.80, draw=pidorange!60!black, line width=0.7pt] coordinates {};
}

\newcommand{\BoxPairHistmRmse}{%
  % HLL
  \addplot+[boxplot prepared={draw position=0.68,
      lower whisker=13.638643, lower quartile=18.919846,
      median=25.535442, upper quartile=28.188578, upper whisker=32.915333},
    fill=pidblue, fill opacity=0.80, draw=pidblue!60!black, line width=0.7pt] coordinates {};
  \addplot+[boxplot prepared={draw position=1.32,
      lower whisker=10.192244, lower quartile=15.124520,
      median=22.660110, upper quartile=25.666630, upper whisker=32.848003},
    fill=pidorange, fill opacity=0.80, draw=pidorange!60!black, line width=0.7pt] coordinates {};
  % HSG
  \addplot+[boxplot prepared={draw position=3.68,
      lower whisker=8.861112, lower quartile=14.823628,
      median=22.874901, upper quartile=25.292922, upper whisker=31.729350},
    fill=pidblue, fill opacity=0.80, draw=pidblue!60!black, line width=0.7pt] coordinates {};
  \addplot+[boxplot prepared={draw position=4.32,
      lower whisker=6.988239,  lower quartile=12.680175,
      median=20.150979, upper quartile=24.699633, upper whisker=31.663811},
    fill=pidorange, fill opacity=0.80, draw=pidorange!60!black, line width=0.7pt] coordinates {};
  % JSS
  \addplot+[boxplot prepared={draw position=6.68,
      lower whisker=7.883414, lower quartile=13.857346,
      median=22.382468, upper quartile=25.660658, upper whisker=39.061157},
    fill=pidblue, fill opacity=0.80, draw=pidblue!60!black, line width=0.7pt] coordinates {};
  \addplot+[boxplot prepared={draw position=7.32,
      lower whisker=6.463932,  lower quartile=12.147326,
      median=20.442896, upper quartile=25.122595, upper whisker=38.991292},
    fill=pidorange, fill opacity=0.80, draw=pidorange!60!black, line width=0.7pt] coordinates {};
  % JSL
  \addplot+[boxplot prepared={draw position=9.68,
      lower whisker=7.864005, lower quartile=13.809835,
      median=22.378066, upper quartile=25.660658, upper whisker=39.061157},
    fill=pidblue, fill opacity=0.80, draw=pidblue!60!black, line width=0.7pt] coordinates {};
  \addplot+[boxplot prepared={draw position=10.32,
      lower whisker=6.446307,  lower quartile=12.147326,
      median=20.442182, upper quartile=25.122595, upper whisker=38.991292},
    fill=pidorange, fill opacity=0.80, draw=pidorange!60!black, line width=0.7pt] coordinates {};
  % JSR
  \addplot+[boxplot prepared={draw position=12.68,
      lower whisker=7.710040, lower quartile=13.155061,
      median=22.039423, upper quartile=25.202357, upper whisker=32.524657},
    fill=pidblue, fill opacity=0.80, draw=pidblue!60!black, line width=0.7pt] coordinates {};
  \addplot+[boxplot prepared={draw position=13.32,
      lower whisker=6.312231,  lower quartile=11.635416,
      median=19.855275, upper quartile=24.775497, upper whisker=32.459914},
    fill=pidorange, fill opacity=0.80, draw=pidorange!60!black, line width=0.7pt] coordinates {};
}

\newcommand{\BoxPairNestedMae}{%
  % HLL
  \addplot+[boxplot prepared={draw position=0.68,
      lower whisker=17.997813, lower quartile=22.853329,
      median=23.974499, upper quartile=28.440020, upper whisker=32.116580},
    fill=pidblue, fill opacity=0.80, draw=pidblue!60!black, line width=0.7pt] coordinates {};
  \addplot+[boxplot prepared={draw position=1.32,
      lower whisker=8.116355,  lower quartile=14.512881,
      median=16.782219, upper quartile=19.618786, upper whisker=22.843678},
    fill=pidorange, fill opacity=0.80, draw=pidorange!60!black, line width=0.7pt] coordinates {};
  % HSG
  \addplot+[boxplot prepared={draw position=3.68,
      lower whisker=12.003774, lower quartile=15.989067,
      median=19.719977, upper quartile=21.628128, upper whisker=24.923985},
    fill=pidblue, fill opacity=0.80, draw=pidblue!60!black, line width=0.7pt] coordinates {};
  \addplot+[boxplot prepared={draw position=4.32,
      lower whisker=6.204995,  lower quartile=10.985955,
      median=13.798145, upper quartile=17.859176, upper whisker=24.588877},
    fill=pidorange, fill opacity=0.80, draw=pidorange!60!black, line width=0.7pt] coordinates {};
  % JSS
  \addplot+[boxplot prepared={draw position=6.68,
      lower whisker=6.744482, lower quartile=11.444889,
      median=16.825301, upper quartile=20.185108, upper whisker=28.082751},
    fill=pidblue, fill opacity=0.80, draw=pidblue!60!black, line width=0.7pt] coordinates {};
  \addplot+[boxplot prepared={draw position=7.32,
      lower whisker=4.961970,  lower quartile=8.630403,
      median=12.070857, upper quartile=17.314061, upper whisker=27.733408},
    fill=pidorange, fill opacity=0.80, draw=pidorange!60!black, line width=0.7pt] coordinates {};
  % JSL
  \addplot+[boxplot prepared={draw position=9.68,
      lower whisker=6.721765, lower quartile=11.419337,
      median=16.818622, upper quartile=20.075132, upper whisker=28.082751},
    fill=pidblue, fill opacity=0.80, draw=pidblue!60!black, line width=0.7pt] coordinates {};
  \addplot+[boxplot prepared={draw position=10.32,
      lower whisker=4.961970,  lower quartile=8.607670,
      median=12.069630, upper quartile=17.279351, upper whisker=27.733408},
    fill=pidorange, fill opacity=0.80, draw=pidorange!60!black, line width=0.7pt] coordinates {};
  % JSR
  \addplot+[boxplot prepared={draw position=12.68,
      lower whisker=6.501176, lower quartile=11.044629,
      median=16.437309, upper quartile=19.032433, upper whisker=24.927267},
    fill=pidblue, fill opacity=0.80, draw=pidblue!60!black, line width=0.7pt] coordinates {};
  \addplot+[boxplot prepared={draw position=13.32,
      lower whisker=4.653559,  lower quartile=8.147210,
      median=12.060480, upper quartile=16.978752, upper whisker=24.588389},
    fill=pidorange, fill opacity=0.80, draw=pidorange!60!black, line width=0.7pt] coordinates {};
}

\newcommand{\BoxPairNestedRmse}{%
  % HLL
  \addplot+[boxplot prepared={draw position=0.68,
      lower whisker=20.113492, lower quartile=26.664642,
      median=29.410511, upper quartile=33.128403, upper whisker=37.466844},
    fill=pidblue, fill opacity=0.80, draw=pidblue!60!black, line width=0.7pt] coordinates {};
  \addplot+[boxplot prepared={draw position=1.32,
      lower whisker=10.253445, lower quartile=18.020492,
      median=21.337616, upper quartile=25.597169, upper whisker=26.802805},
    fill=pidorange, fill opacity=0.80, draw=pidorange!60!black, line width=0.7pt] coordinates {};
  % HSG
  \addplot+[boxplot prepared={draw position=3.68,
      lower whisker=14.267532, lower quartile=19.459263,
      median=26.047175, upper quartile=27.522777, upper whisker=33.307448},
    fill=pidblue, fill opacity=0.80, draw=pidblue!60!black, line width=0.7pt] coordinates {};
  \addplot+[boxplot prepared={draw position=4.32,
      lower whisker=8.054867,  lower quartile=13.473572,
      median=19.259842, upper quartile=23.847941, upper whisker=32.867130},
    fill=pidorange, fill opacity=0.80, draw=pidorange!60!black, line width=0.7pt] coordinates {};
  % JSS
  \addplot+[boxplot prepared={draw position=6.68,
      lower whisker=9.077457, lower quartile=15.051029,
      median=23.883453, upper quartile=27.010560, upper whisker=41.689632},
    fill=pidblue, fill opacity=0.80, draw=pidblue!60!black, line width=0.7pt] coordinates {};
  \addplot+[boxplot prepared={draw position=7.32,
      lower whisker=7.025720,  lower quartile=12.165637,
      median=17.677670, upper quartile=23.654703, upper whisker=25.988358},
    fill=pidorange, fill opacity=0.80, draw=pidorange!60!black, line width=0.7pt] coordinates {};
  % JSL
  \addplot+[boxplot prepared={draw position=9.68,
      lower whisker=9.077323, lower quartile=15.006005,
      median=23.878928, upper quartile=26.956742, upper whisker=41.689632},
    fill=pidblue, fill opacity=0.80, draw=pidblue!60!black, line width=0.7pt] coordinates {};
  \addplot+[boxplot prepared={draw position=10.32,
      lower whisker=7.025720,  lower quartile=12.124142,
      median=17.677192, upper quartile=23.654703, upper whisker=25.988358},
    fill=pidorange, fill opacity=0.80, draw=pidorange!60!black, line width=0.7pt] coordinates {};
  % JSR
  \addplot+[boxplot prepared={draw position=12.68,
      lower whisker=9.070546, lower quartile=14.395533,
      median=23.483599, upper quartile=25.976735, upper whisker=34.055810},
    fill=pidblue, fill opacity=0.80, draw=pidblue!60!black, line width=0.7pt] coordinates {};
  \addplot+[boxplot prepared={draw position=13.32,
      lower whisker=6.568299,  lower quartile=11.353430,
      median=17.672535, upper quartile=23.364839, upper whisker=33.601741},
    fill=pidorange, fill opacity=0.80, draw=pidorange!60!black, line width=0.7pt] coordinates {};
}

% ── Figure ──────────────────────────────────────────────────
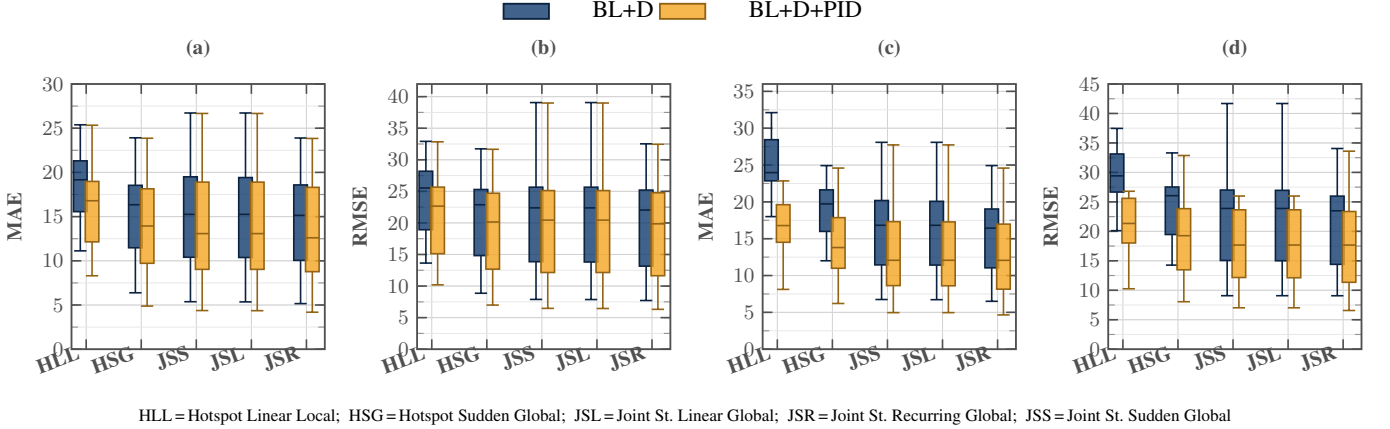
\begin{figure*}[t]
\centering
%
% Legend strip ─────────────────────────────────────────────
\begin{tikzpicture}
  \begin{axis}[
    hide axis,
    xmin=0, xmax=1, ymin=0, ymax=1,
    width=\textwidth, height=2.45cm,
    legend columns=2,
    legend style={
      draw=none,
      fill=none,
      at={(0.5,0.5)},
      anchor=center,
      column sep=1.4em,
      /tikz/every even column/.append style={column sep=0pt},
      font=\small,
    },
  ]
    \addlegendimage{fill=pidblue,   fill opacity=0.80, draw=pidblue!60!black,
                    area legend, line width=0.7pt}
    \addlegendentry{BL+D}
    \addlegendimage{fill=pidorange, fill opacity=0.80, draw=pidorange!60!black,
                    area legend, line width=0.7pt}
    \addlegendentry{BL+D+PID}
  \end{axis}
\end{tikzpicture}
\vspace{2pt}
%
% Four sub-plots ────────────────────────────────────────────
\resizebox{\textwidth}{!}{%
\begin{tikzpicture}
\begin{groupplot}[
  % ── layout ──
  group style={
    group size=4 by 1,
    horizontal sep=4em,          % generous gap so y-axes breathe
  },
  width=0.30\textwidth,
  height=0.31\textwidth,
  % ── box-plot direction ──
  boxplot/draw direction=y,
  % ── x axis ──
  xtick={1,4,7,10,13},
  xticklabels={HLL,HSG,JSS,JSL,JSR},
  xmin=0.2, xmax=14.0,
  x tick label style={
    rotate=20,
    anchor=east,
    font=\small\bfseries,
    color=axgray,
  },
  % ── y axis (shared defaults; overridden per panel) ──
  ytick distance=5,
  minor ytick={0,2.5,5,...,50},  % half-step minor ticks
  yminorticks=true,
  yticklabel style={
    font=\small,
    color=axgray,
    /pgf/number format/fixed,
    /pgf/number format/precision=0,
  },
  ylabel style={font=\small\bfseries, color=axgray, yshift=2pt},
  % ── grid ──
  grid=both,
  grid style={gridgray, line width=0.4pt},
  minor grid style={gridgray!50, line width=0.25pt},
  major grid style={gridgray, line width=0.4pt},
  % ── frame ──
  axis line style={axgray, line width=0.8pt},
  tick style={color=axgray, line width=0.7pt},
  minor tick style={color=axgray!60, line width=0.5pt},
  % ── box style (defaults) ──
  every boxplot/.style={line width=0.9pt, solid},
  every boxplot box/.style={solid, line width=0.9pt},
  every boxplot whisker/.style={solid, line width=0.9pt, densely dashed},
  every boxplot median/.style={solid, line width=1.4pt, color=white},
  % ── title ──
  title style={
    font=\small\bfseries,
    color=axgray,
    align=center,
    yshift=2pt,
  },
  % ── box width ──
  boxplot/box extend=0.74,
  boxplot/whisker extend=0.67,
]

% ── Panel (a): HiSTM MAE ─────────────────────────────────
\nextgroupplot[
  title={(a)},
  ylabel={MAE},
  ymin=0, ymax=30,
  ytick={0,5,10,15,20,25,30},
]
\BoxPairHistmMae

% ── Panel (b): HiSTM RMSE ────────────────────────────────
\nextgroupplot[
  title={(b)},
  ylabel={RMSE},
  ymin=0, ymax=42,
  ytick={0,5,10,15,20,25,30,35,40},
]
\BoxPairHistmRmse

% ── Panel (c): HiSTM\_nested MAE ─────────────────────────
\nextgroupplot[
  title={(c)},
  ylabel={MAE},
  ymin=0, ymax=36,
  ytick={0,5,10,15,20,25,30,35},
]
\BoxPairNestedMae

% ── Panel (d): HiSTM\_nested RMSE ────────────────────────
\nextgroupplot[
  title={(d)},
  ylabel={RMSE},
  ymin=0, ymax=45,
  ytick={0,5,10,15,20,25,30,35,40,45},
]
\BoxPairNestedRmse

\end{groupplot}
\end{tikzpicture}
}% end resizebox
\vspace{0.0em}
{\fontsize{6.5}{7.5}\selectfont
HLL\,=\,Hotspot Linear Local;\enspace
HSG\,=\,Hotspot Sudden Global;\enspace
JSL\,=\,Joint St.\ Linear Global;\enspace
JSR\,=\,Joint St.\ Recurring Global;\enspace
JSS\,=\,Joint St.\ Sudden Global}

\caption{%
Box plots comparing error distributions with and without PID correction across five drift scenarios for both HiSTM and HiSTM\_Nested. (a) HiSTM MAE. (b) HiSTM RMSE. (c) HiSTM\_Nested MAE. (d) HiSTM\_Nested RMSE.
}
\label{fig:pid_boxplots_four_panel}
\end{figure*}

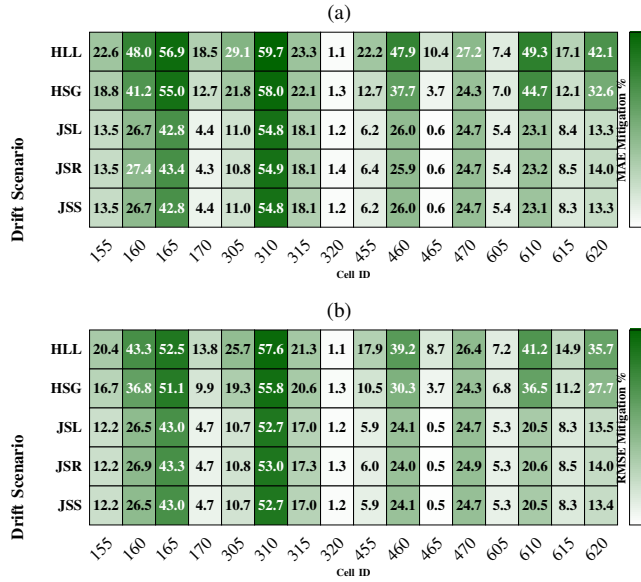
\begin{figure}[htbp]
\centering

% =========================================================
%  SHARED MACROS
% =========================================================
\def\cw{0.55cm}
\def\ch{0.65cm}

\newcommand{\heatcell}[5]{%
  \pgfmathsetmacro{\ratio}{min(#3/#4,1.0)}%
  \pgfmathsetmacro{\rr}{1 - \ratio}%
  \pgfmathsetmacro{\gg}{1 - 0.608*\ratio}%
  \pgfmathsetmacro{\bb}{1 - \ratio}%
  \definecolor{cellcolor}{rgb}{\rr,\gg,\bb}%
  \pgfmathsetmacro{\textwhite}{(\ratio > 0.45) ? 1 : 0}%
  \ifdim\textwhite pt>0.5pt \def\tcol{white}\else\def\tcol{black}\fi
  \fill[cellcolor]
    (#1*\cw,#2*\ch) rectangle (#1*\cw+\cw,#2*\ch+\ch);
  \draw[black,line width=0.2pt]
    (#1*\cw,#2*\ch) rectangle (#1*\cw+\cw,#2*\ch+\ch);
  \node[\tcol,font=\fontsize{7}{7}\selectfont\bfseries] at
  %\node[\tcol,font=\fontsize{4}{4}\selectfont\bfseries] at
    (#1*\cw+0.5*\cw,#2*\ch+0.5*\ch) {#5};
}

\newcommand{\drawcolorbar}[3]{%
  \def\cbx{#1}\def\cbh{#2}\def\cblabel{#3}%
  \def\cbw{0.20cm}%
  \foreach \k in {0,...,99}{%
    \pgfmathsetmacro{\yr}{\k/100}%
    \pgfmathsetmacro{\rr}{1-\yr}%
    \pgfmathsetmacro{\gg}{1-0.608*\yr}%
    \pgfmathsetmacro{\bb}{1-\yr}%
    \definecolor{cbcol}{rgb}{\rr,\gg,\bb}%
    \fill[cbcol](\cbx,\yr*\cbh) rectangle (\cbx+\cbw,\yr*\cbh+\cbh/100+0.02cm);%
  }%
  \draw[black,line width=0.3pt](\cbx,0) rectangle (\cbx+\cbw,\cbh);%
  \foreach \v in {0,10,20,30}{%
    \pgfmathsetmacro{\yr}{\v/60}%
    \draw[black,line width=0.3pt](\cbx+\cbw,\yr*\cbh)--(\cbx+\cbw+0.07cm,\yr*\cbh);%
    \node[right,font=\fontsize{6}{6}\selectfont] at (\cbx+\cbw+0.09cm,\yr*\cbh) {\v};%
  }%
  \node[rotate=90,font=\fontsize{6}{6}\selectfont\bfseries,anchor=south]
    at (\cbx+0.5*\cbw,0.5*\cbh) {\cblabel};%
}

% =========================================================
%  (a)  MAE Mitigation
% =========================================================

{\fontsize{8}{9}\selectfont(a)}\\[0.3em]
\resizebox{\columnwidth}{!}{%
\begin{tikzpicture}[font=\fontsize{4}{4}\selectfont]

% Row 4: Hotspot Linear
\heatcell{0}{4}{22.6}{60}{22.6} \heatcell{1}{4}{48.0}{60}{48.0}
\heatcell{2}{4}{56.9}{60}{56.9} \heatcell{3}{4}{18.5}{60}{18.5}
\heatcell{4}{4}{29.1}{60}{29.1} \heatcell{5}{4}{59.7}{60}{59.7}
\heatcell{6}{4}{23.3}{60}{23.3} \heatcell{7}{4}{1.1}{60}{1.1}
\heatcell{8}{4}{22.2}{60}{22.2} \heatcell{9}{4}{47.9}{60}{47.9}
\heatcell{10}{4}{10.4}{60}{10.4}\heatcell{11}{4}{27.2}{60}{27.2}
\heatcell{12}{4}{7.4}{60}{7.4}  \heatcell{13}{4}{49.3}{60}{49.3}
\heatcell{14}{4}{17.1}{60}{17.1}\heatcell{15}{4}{42.1}{60}{42.1}
% Row 3: Hotspot Sudden
\heatcell{0}{3}{18.8}{60}{18.8} \heatcell{1}{3}{41.2}{60}{41.2}
\heatcell{2}{3}{55.0}{60}{55.0} \heatcell{3}{3}{12.7}{60}{12.7}
\heatcell{4}{3}{21.8}{60}{21.8} \heatcell{5}{3}{58.0}{60}{58.0}
\heatcell{6}{3}{22.1}{60}{22.1} \heatcell{7}{3}{1.3}{60}{1.3}
\heatcell{8}{3}{12.7}{60}{12.7} \heatcell{9}{3}{37.7}{60}{37.7}
\heatcell{10}{3}{3.7}{60}{3.7}  \heatcell{11}{3}{24.3}{60}{24.3}
\heatcell{12}{3}{7.0}{60}{7.0}  \heatcell{13}{3}{44.7}{60}{44.7}
\heatcell{14}{3}{12.1}{60}{12.1}\heatcell{15}{3}{32.6}{60}{32.6}
% Row 2: Joint St Linear
\heatcell{0}{2}{13.5}{60}{13.5} \heatcell{1}{2}{26.7}{60}{26.7}
\heatcell{2}{2}{42.8}{60}{42.8} \heatcell{3}{2}{4.4}{60}{4.4}
\heatcell{4}{2}{11.0}{60}{11.0} \heatcell{5}{2}{54.8}{60}{54.8}
\heatcell{6}{2}{18.1}{60}{18.1} \heatcell{7}{2}{1.2}{60}{1.2}
\heatcell{8}{2}{6.2}{60}{6.2}   \heatcell{9}{2}{26.0}{60}{26.0}
\heatcell{10}{2}{0.6}{60}{0.6}  \heatcell{11}{2}{24.7}{60}{24.7}
\heatcell{12}{2}{5.4}{60}{5.4}  \heatcell{13}{2}{23.1}{60}{23.1}
\heatcell{14}{2}{8.4}{60}{8.4}  \heatcell{15}{2}{13.3}{60}{13.3}
% Row 1: Joint St Recurring
\heatcell{0}{1}{13.5}{60}{13.5} \heatcell{1}{1}{27.4}{60}{27.4}
\heatcell{2}{1}{43.4}{60}{43.4} \heatcell{3}{1}{4.3}{60}{4.3}
\heatcell{4}{1}{10.8}{60}{10.8} \heatcell{5}{1}{54.9}{60}{54.9}
\heatcell{6}{1}{18.1}{60}{18.1} \heatcell{7}{1}{1.4}{60}{1.4}
\heatcell{8}{1}{6.4}{60}{6.4}   \heatcell{9}{1}{25.9}{60}{25.9}
\heatcell{10}{1}{0.6}{60}{0.6}  \heatcell{11}{1}{24.7}{60}{24.7}
\heatcell{12}{1}{5.4}{60}{5.4}  \heatcell{13}{1}{23.2}{60}{23.2}
\heatcell{14}{1}{8.5}{60}{8.5}  \heatcell{15}{1}{14.0}{60}{14.0}
% Row 0: Joint St Sudden
\heatcell{0}{0}{13.5}{60}{13.5} \heatcell{1}{0}{26.7}{60}{26.7}
\heatcell{2}{0}{42.8}{60}{42.8} \heatcell{3}{0}{4.4}{60}{4.4}
\heatcell{4}{0}{11.0}{60}{11.0} \heatcell{5}{0}{54.8}{60}{54.8}
\heatcell{6}{0}{18.1}{60}{18.1} \heatcell{7}{0}{1.2}{60}{1.2}
\heatcell{8}{0}{6.2}{60}{6.2}   \heatcell{9}{0}{26.0}{60}{26.0}
\heatcell{10}{0}{0.6}{60}{0.6}  \heatcell{11}{0}{24.7}{60}{24.7}
\heatcell{12}{0}{5.4}{60}{5.4}  \heatcell{13}{0}{23.1}{60}{23.1}
\heatcell{14}{0}{8.3}{60}{8.3}  \heatcell{15}{0}{13.3}{60}{13.3}

% x-axis labels
\foreach \i/\lbl in {
  0/155,1/160,2/165,3/170,4/305,5/310,6/315,7/320,
  8/455,9/460,10/465,11/470,12/605,13/610,14/615,15/620}{
  \node[below,font=\fontsize{8}{8}\selectfont,rotate=45,anchor=north east]
    at (\i*\cw+0.5*\cw,0) {\lbl};
}
\node[below=0.6cm,font=\fontsize{5}{5}\selectfont\bfseries] at (8*\cw,0) {Cell ID};

\foreach \j/\lbl in {
  0/{JSS},1/{JSR},2/{JSL},3/{HSG},4/{HLL}}{
  \node[left,font=\fontsize{7}{8}\selectfont\bfseries,align=right]
    at (0,\j*\ch+0.5*\ch) {\lbl};
}
% y-axis title
\node[left=1.2cm,rotate=90,font=\fontsize{8}{8}\selectfont\bfseries]
  at (0,2.5*\ch) {Drift Scenario};

% colorbar
\drawcolorbar{16*\cw+0.20cm}{5*\ch}{MAE Mitigation \%}

\end{tikzpicture}%
}% end resizebox

% =========================================================
%  (b)  RMSE Mitigation
% =========================================================
% CHANGED: smaller title font
{\fontsize{8}{9}\selectfont(b)}\\[0.3em]
\resizebox{\columnwidth}{!}{%
\begin{tikzpicture}[font=\fontsize{4}{4}\selectfont]

% Row 4: Hotspot Linear
\heatcell{0}{4}{20.4}{60}{20.4} \heatcell{1}{4}{43.3}{60}{43.3}
\heatcell{2}{4}{52.5}{60}{52.5} \heatcell{3}{4}{13.8}{60}{13.8}
\heatcell{4}{4}{25.7}{60}{25.7} \heatcell{5}{4}{57.6}{60}{57.6}
\heatcell{6}{4}{21.3}{60}{21.3} \heatcell{7}{4}{1.1}{60}{1.1}
\heatcell{8}{4}{17.9}{60}{17.9} \heatcell{9}{4}{39.2}{60}{39.2}
\heatcell{10}{4}{8.7}{60}{8.7}  \heatcell{11}{4}{26.4}{60}{26.4}
\heatcell{12}{4}{7.2}{60}{7.2}  \heatcell{13}{4}{41.2}{60}{41.2}
\heatcell{14}{4}{14.9}{60}{14.9}\heatcell{15}{4}{35.7}{60}{35.7}
% Row 3: Hotspot Sudden
\heatcell{0}{3}{16.7}{60}{16.7} \heatcell{1}{3}{36.8}{60}{36.8}
\heatcell{2}{3}{51.1}{60}{51.1} \heatcell{3}{3}{9.9}{60}{9.9}
\heatcell{4}{3}{19.3}{60}{19.3} \heatcell{5}{3}{55.8}{60}{55.8}
\heatcell{6}{3}{20.6}{60}{20.6} \heatcell{7}{3}{1.3}{60}{1.3}
\heatcell{8}{3}{10.5}{60}{10.5} \heatcell{9}{3}{30.3}{60}{30.3}
\heatcell{10}{3}{3.7}{60}{3.7}  \heatcell{11}{3}{24.3}{60}{24.3}
\heatcell{12}{3}{6.8}{60}{6.8}  \heatcell{13}{3}{36.5}{60}{36.5}
\heatcell{14}{3}{11.2}{60}{11.2}\heatcell{15}{3}{27.7}{60}{27.7}
% Row 2: Joint St Linear
\heatcell{0}{2}{12.2}{60}{12.2} \heatcell{1}{2}{26.5}{60}{26.5}
\heatcell{2}{2}{43.0}{60}{43.0} \heatcell{3}{2}{4.7}{60}{4.7}
\heatcell{4}{2}{10.7}{60}{10.7} \heatcell{5}{2}{52.7}{60}{52.7}
\heatcell{6}{2}{17.0}{60}{17.0} \heatcell{7}{2}{1.2}{60}{1.2}
\heatcell{8}{2}{5.9}{60}{5.9}   \heatcell{9}{2}{24.1}{60}{24.1}
\heatcell{10}{2}{0.5}{60}{0.5}  \heatcell{11}{2}{24.7}{60}{24.7}
\heatcell{12}{2}{5.3}{60}{5.3}  \heatcell{13}{2}{20.5}{60}{20.5}
\heatcell{14}{2}{8.3}{60}{8.3}  \heatcell{15}{2}{13.5}{60}{13.5}
% Row 1: Joint St Recurring
\heatcell{0}{1}{12.2}{60}{12.2} \heatcell{1}{1}{26.9}{60}{26.9}
\heatcell{2}{1}{43.3}{60}{43.3} \heatcell{3}{1}{4.7}{60}{4.7}
\heatcell{4}{1}{10.8}{60}{10.8} \heatcell{5}{1}{53.0}{60}{53.0}
\heatcell{6}{1}{17.3}{60}{17.3} \heatcell{7}{1}{1.3}{60}{1.3}
\heatcell{8}{1}{6.0}{60}{6.0}   \heatcell{9}{1}{24.0}{60}{24.0}
\heatcell{10}{1}{0.5}{60}{0.5}  \heatcell{11}{1}{24.9}{60}{24.9}
\heatcell{12}{1}{5.3}{60}{5.3}  \heatcell{13}{1}{20.6}{60}{20.6}
\heatcell{14}{1}{8.5}{60}{8.5}  \heatcell{15}{1}{14.0}{60}{14.0}
% Row 0: Joint St Sudden
\heatcell{0}{0}{12.2}{60}{12.2} \heatcell{1}{0}{26.5}{60}{26.5}
\heatcell{2}{0}{43.0}{60}{43.0} \heatcell{3}{0}{4.7}{60}{4.7}
\heatcell{4}{0}{10.7}{60}{10.7} \heatcell{5}{0}{52.7}{60}{52.7}
\heatcell{6}{0}{17.0}{60}{17.0} \heatcell{7}{0}{1.2}{60}{1.2}
\heatcell{8}{0}{5.9}{60}{5.9}   \heatcell{9}{0}{24.1}{60}{24.1}
\heatcell{10}{0}{0.5}{60}{0.5}  \heatcell{11}{0}{24.7}{60}{24.7}
\heatcell{12}{0}{5.3}{60}{5.3}  \heatcell{13}{0}{20.5}{60}{20.5}
\heatcell{14}{0}{8.3}{60}{8.3}  \heatcell{15}{0}{13.4}{60}{13.4}

% x-axis labels
\foreach \i/\lbl in {
  0/155,1/160,2/165,3/170,4/305,5/310,6/315,7/320,
  8/455,9/460,10/465,11/470,12/605,13/610,14/615,15/620}{
  \node[below,font=\fontsize{8}{8}\selectfont,rotate=45,anchor=north east]
    at (\i*\cw+0.5*\cw,0) {\lbl};
}
\node[below=0.6cm,font=\fontsize{5}{5}\selectfont\bfseries] at (8*\cw,0) {Cell ID};

\foreach \j/\lbl in {
  0/{JSS},1/{JSR},2/{JSL},3/{HSG},4/{HLL}}{
  \node[left,font=\fontsize{7}{8}\selectfont\bfseries,align=right]
    at (0,\j*\ch+0.5*\ch) {\lbl};
}
% y-axis title
\node[left=1.2cm,rotate=90,font=\fontsize{8}{8}\selectfont\bfseries]
  at (0,2.5*\ch) {Drift Scenario};

% colorbar
\drawcolorbar{16*\cw+0.20cm}{5*\ch}{RMSE Mitigation \%}

\end{tikzpicture}%
}

\caption{Heatmaps of selected cells depicting mitigation percentage under drift with PID correction for the HiSTM\_Nested model. (a) MAE Mitigation \%. (b) RMSE Mitigation \%.}
\label{fig:mitigation_heatmaps}
\end{figure}

% -------------------------------------------------------
%  Required packages:
%    \usepackage{tikz}
%    \usepackage{xcolor}
%    \usepackage{pgfmath}
% -------------------------------------------------------

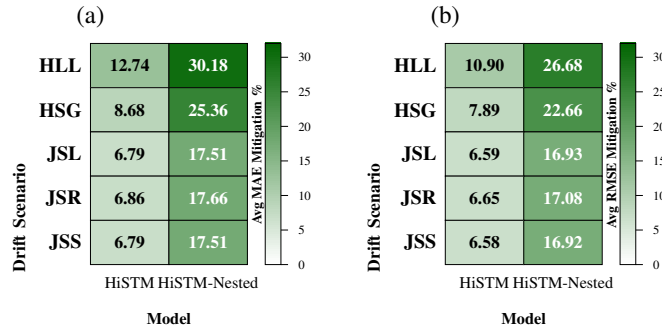
\begin{figure}[htbp]
\centering

% =========================================================
%  SHARED MACROS
% =========================================================
\def\cw{1.10cm}
\def\ch{0.62cm}

\newcommand{\heatcellwide}[5]{%
  \pgfmathsetmacro{\ratio}{min(#3/#4,1.0)}%
  \pgfmathsetmacro{\rr}{1 - \ratio}%
  \pgfmathsetmacro{\gg}{1 - 0.608*\ratio}%
  \pgfmathsetmacro{\bb}{1 - \ratio}%
  \definecolor{cellcolor}{rgb}{\rr,\gg,\bb}%
  \pgfmathsetmacro{\textwhite}{(\ratio > 0.45) ? 1 : 0}%
  \ifdim\textwhite pt>0.5pt \def\tcol{white}\else\def\tcol{black}\fi
  \fill[cellcolor]
    (#1*\cw,#2*\ch) rectangle (#1*\cw+\cw,#2*\ch+\ch);
  \draw[black,line width=0.3pt]
    (#1*\cw,#2*\ch) rectangle (#1*\cw+\cw,#2*\ch+\ch);
  \node[\tcol,font=\fontsize{7}{7}\selectfont\bfseries] at
    (#1*\cw+0.5*\cw,#2*\ch+0.5*\ch) {#5};
}

\newcommand{\drawcolorbarvert}[3]{%
  \def\cbx{#1}\def\cbh{#2}\def\cblabel{#3}%
  \def\cbw{0.22cm}%
  \foreach \k in {0,...,99}{%
    \pgfmathsetmacro{\yr}{\k/100}%
    \pgfmathsetmacro{\rr}{1-\yr}%
    \pgfmathsetmacro{\gg}{1-0.608*\yr}%
    \pgfmathsetmacro{\bb}{1-\yr}%
    \definecolor{cbcol}{rgb}{\rr,\gg,\bb}%
    \fill[cbcol](\cbx,\yr*\cbh) rectangle (\cbx+\cbw,\yr*\cbh+\cbh/100+0.02cm);%
  }%
  \draw[black,line width=0.3pt](\cbx,0) rectangle (\cbx+\cbw,\cbh);%
  \foreach \v in {0,5,10,15,20,25,30}{%
    \pgfmathsetmacro{\yr}{\v/32}%
    \draw[black,line width=0.3pt](\cbx+\cbw,\yr*\cbh)--(\cbx+\cbw+0.10cm,\yr*\cbh);%
    \node[right,font=\fontsize{5}{5}\selectfont] at (\cbx+\cbw+0.12cm,\yr*\cbh) {\v};%
  }%
  \node[rotate=90,font=\fontsize{5.5}{5.5}\selectfont\bfseries,anchor=south]
    at (\cbx+0.5*\cbw,0.5*\cbh) {\cblabel};%
}

% =========================================================
%  LEFT HEATMAP: (a) Average MAE Mitigation %
% =========================================================
\begin{minipage}[t]{0.47\columnwidth}
\centering
\resizebox{\linewidth}{!}{%
\begin{tikzpicture}[font=\fontsize{6}{6}\selectfont]

% Row 4: Hotspot Linear Local
\heatcellwide{0}{4}{12.74}{32}{12.74}
\heatcellwide{1}{4}{30.17}{32}{30.18}
% Row 3: Hotspot Sudden Global
\heatcellwide{0}{3}{8.68}{32}{8.68}
\heatcellwide{1}{3}{25.36}{32}{25.36}
% Row 2: Joint St Linear Global
\heatcellwide{0}{2}{6.79}{32}{6.79}
\heatcellwide{1}{2}{17.50}{32}{17.51}
% Row 1: Joint St Recurring Global
\heatcellwide{0}{1}{6.86}{32}{6.86}
\heatcellwide{1}{1}{17.65}{32}{17.66}
% Row 0: Joint St Sudden Global
\heatcellwide{0}{0}{6.79}{32}{6.79}
\heatcellwide{1}{0}{17.50}{32}{17.51}

% (a) label above heatmap
\node[font=\fontsize{10}{11}\selectfont,anchor=south]
  at (0,5*\ch+0.10cm) {(a)};

% x-axis labels
\foreach \i/\lbl in {0/{HiSTM},1/{HiSTM-Nested}}{
  \node[below,font=\fontsize{6.5}{7}\selectfont,align=center]
    at (\i*\cw+0.5*\cw,-0.05cm) {\lbl};
}
\node[below=0.55cm,font=\fontsize{6.5}{7}\selectfont\bfseries]
  at (1*\cw,0) {Model};

% CHANGED: abbreviated y-axis labels, larger font
\foreach \j/\lbl in {
  0/{JSS},1/{JSR},2/{JSL},3/{HSG},4/{HLL}}{
  \node[left,font=\fontsize{8}{9}\selectfont\bfseries,align=right]
    at (0,\j*\ch+0.5*\ch) {\lbl};
}
% y-axis title
\node[left=1.0cm,rotate=90,font=\fontsize{7}{7}\selectfont\bfseries]
  at (0,2.5*\ch) {Drift Scenario};

% colorbar
\drawcolorbarvert{2*\cw+0.25cm}{5*\ch}{Avg MAE Mitigation \%}

\end{tikzpicture}%
}
\end{minipage}%
\hfill
% =========================================================
%  RIGHT HEATMAP: (b) Average RMSE Mitigation %
% =========================================================
\begin{minipage}[t]{0.47\columnwidth}
\centering
\resizebox{\linewidth}{!}{%
\begin{tikzpicture}[font=\fontsize{6}{6}\selectfont]

% Row 4: Hotspot Linear Local
\heatcellwide{0}{4}{10.90}{32}{10.90}
\heatcellwide{1}{4}{26.68}{32}{26.68}
% Row 3: Hotspot Sudden Global
\heatcellwide{0}{3}{7.89}{32}{7.89}
\heatcellwide{1}{3}{22.66}{32}{22.66}
% Row 2: Joint St Linear Global
\heatcellwide{0}{2}{6.59}{32}{6.59}
\heatcellwide{1}{2}{16.94}{32}{16.93}
% Row 1: Joint St Recurring Global
\heatcellwide{0}{1}{6.65}{32}{6.65}
\heatcellwide{1}{1}{17.08}{32}{17.08}
% Row 0: Joint St Sudden Global
\heatcellwide{0}{0}{6.58}{32}{6.58}
\heatcellwide{1}{0}{16.93}{32}{16.92}

% (b) label above heatmap
\node[font=\fontsize{10}{11}\selectfont,anchor=south]
  at (0,5*\ch+0.10cm) {(b)};

% x-axis labels
\foreach \i/\lbl in {0/{HiSTM},1/{HiSTM-Nested}}{
  \node[below,font=\fontsize{6.5}{7}\selectfont,align=center]
    at (\i*\cw+0.5*\cw,-0.05cm) {\lbl};
}
\node[below=0.55cm,font=\fontsize{6.5}{7}\selectfont\bfseries]
  at (1*\cw,0) {Model};

% CHANGED: abbreviated y-axis labels, larger font
\foreach \j/\lbl in {
  0/{JSS},1/{JSR},2/{JSL},3/{HSG},4/{HLL}}{
  \node[left,font=\fontsize{8}{9}\selectfont\bfseries,align=right]
    at (0,\j*\ch+0.5*\ch) {\lbl};
}
% y-axis title
\node[left=1.0cm,rotate=90,font=\fontsize{7}{7}\selectfont\bfseries]
  at (0,2.5*\ch) {Drift Scenario};

% colorbar
\drawcolorbarvert{2*\cw+0.25cm}{5*\ch}{Avg RMSE Mitigation \%}

\end{tikzpicture}%
}
\end{minipage}

\caption{Average mitigation (\%) of the HiSTM and HiSTM\_Nested models under PID correction, aggregated across 16 selected cells. (a) Average MAE Mitigation \%. (b) Average RMSE Mitigation \%.}
\label{fig:avg_mitigation}
\end{figure}

% \input{images/figure_8}
% \input{images/figure_9}
% \input{images/figure_8_9}
% Required in the preamble:
% \usepackage{pgfplots}
% \usepgfplotslibrary{groupplots}
% \usetikzlibrary{calc}
% \pgfplotsset{compat=1.18}

% Styles and helper macros
% Colorblind-friendly Okabe-Ito inspired palette
\definecolor{CleanBlue}{HTML}{0072B2}
\definecolor{DriftOrange}{HTML}{D55E00}
\definecolor{PidGreen}{HTML}{009E73}
\definecolor{PidPurple}{HTML}{CC79A7}

\pgfplotsset{
    cleanline/.style={
        CleanBlue!55,
        line width=0.95pt
    },
    driftline/.style={
        DriftOrange!65,
        line width=1.00pt
    },
    blpoint/.style={
        only marks,
        mark=*,
        mark size=2.05pt,
        mark options={
            fill=white,
            draw=CleanBlue,
            line width=0.65pt
        }
    },
    blpidpoint/.style={
        only marks,
        mark=*,
        mark size=2.05pt,
        mark options={
            fill=CleanBlue,
            draw=CleanBlue,
            line width=0.65pt
        }
    },
    bdpoint/.style={
        only marks,
        mark=square*,
        mark size=2.15pt,
        mark options={
            fill=white,
            draw=DriftOrange,
            line width=0.65pt
        }
    },
    bdpidpoint/.style={
        only marks,
        mark=square*,
        mark size=2.15pt,
        mark options={
            fill=DriftOrange,
            draw=DriftOrange,
            line width=0.65pt
        }
    }
}
\newcommand{\cleanpair}[3]{%
    \addplot[cleanline] coordinates {(#2,#1) (#3,#1)};
    \addplot[blpoint] coordinates {(#2,#1)};
    \addplot[blpidpoint] coordinates {(#3,#1)};
}

\newcommand{\driftpair}[3]{%
    \addplot[driftline] coordinates {(#2,#1) (#3,#1)};
    \addplot[bdpoint] coordinates {(#2,#1)};
    \addplot[bdpidpoint] coordinates {(#3,#1)};
}

\begin{figure*}[t]
\centering

\begin{tikzpicture}
\begin{groupplot}[
    group style={
        group size=4 by 1,
        horizontal sep=0.3cm,
    },
    width=0.31\textwidth,
    height=6.0cm,
    ymin=0.45,
    ymax=16.55,
    y dir=reverse,
    ytick={1,2,3,4,5,6,7,8,9,10,11,12,13,14,15,16},
    yticklabels={
        155,160,165,170,
        305,310,315,320,
        455,460,465,470,
        605,610,615,620
    },
    xminorticks=false,
    grid=major,
    xmajorgrids=true,
    ymajorgrids=true,
    grid style={black!12, line width=0.25pt},
    axis line style={black!70},
    tick style={black!70},
    tick label style={font=\footnotesize},
    label style={font=\footnotesize},
    title style={font=\footnotesize},
    xlabel style={yshift=0.3ex},
    ylabel style={yshift=-0.3ex},
]

% =========================================================
% (a) MAE - HiSTM
% =========================================================
\nextgroupplot[
    title={(a)},
    xmin=4, xmax=34,
    xtick={5,10,15,20,25,30},
    xlabel={MAE},
    ylabel={Cell ID}
]
\cleanpair{0.82}{12.65}{12.58}
\driftpair{1.18}{18.49}{18.33}
\cleanpair{1.82}{12.73}{12.63}
\driftpair{2.18}{18.35}{16.48}
\cleanpair{2.82}{6.91}{6.67}
\driftpair{3.18}{11.12}{10.93}
\cleanpair{3.82}{15.80}{15.41}
\driftpair{4.18}{21.00}{18.55}
\cleanpair{4.82}{10.88}{9.44}
\driftpair{5.18}{16.65}{12.55}
\cleanpair{5.82}{15.49}{12.29}
\driftpair{6.18}{21.10}{14.18}
\cleanpair{6.82}{19.37}{18.28}
\driftpair{7.18}{22.68}{19.54}
\cleanpair{7.82}{23.33}{23.28}
\driftpair{8.18}{25.39}{25.34}
\cleanpair{8.82}{16.13}{15.96}
\driftpair{9.18}{19.35}{17.12}
\cleanpair{9.82}{14.63}{12.61}
\driftpair{10.18}{18.98}{15.57}
\cleanpair{10.82}{19.49}{19.34}
\driftpair{11.18}{23.97}{21.54}
\cleanpair{11.82}{18.96}{18.23}
\driftpair{12.18}{19.87}{18.79}
\cleanpair{12.82}{6.63}{6.41}
\driftpair{13.18}{11.46}{10.11}
\cleanpair{13.82}{5.09}{4.13}
\driftpair{14.18}{12.00}{8.31}
\cleanpair{14.82}{17.93}{17.86}
\driftpair{15.18}{21.94}{21.45}
\cleanpair{15.82}{7.33}{5.80}
\driftpair{16.18}{12.28}{10.07}

% =========================================================
% (b) RMSE - HiSTM
% =========================================================
\nextgroupplot[
    title={(b)},
    xmin=6, xmax=39,
    xtick={10,15,20,25,30,35},
    xlabel={RMSE},
    yticklabels={}
]
\cleanpair{0.82}{17.01}{16.93}
\driftpair{1.18}{22.67}{22.49}
\cleanpair{1.82}{17.29}{17.15}
\driftpair{2.18}{22.88}{21.08}
\cleanpair{2.82}{10.04}{9.66}
\driftpair{3.18}{13.73}{13.42}
\cleanpair{3.82}{21.92}{21.41}
\driftpair{4.18}{28.09}{25.59}
\cleanpair{4.82}{14.11}{12.25}
\driftpair{5.18}{20.28}{15.69}
\cleanpair{5.82}{22.44}{18.21}
\driftpair{6.18}{27.04}{19.66}
\cleanpair{6.82}{26.14}{24.64}
\driftpair{7.18}{29.25}{25.91}
\cleanpair{7.82}{31.46}{31.39}
\driftpair{8.18}{32.92}{32.85}
\cleanpair{8.82}{22.05}{21.83}
\driftpair{9.18}{24.79}{22.83}
\cleanpair{9.82}{24.35}{21.32}
\driftpair{10.18}{27.04}{23.04}
\cleanpair{10.82}{25.18}{24.98}
\driftpair{11.18}{30.24}{27.67}
\cleanpair{11.82}{26.18}{25.04}
\driftpair{12.18}{26.28}{25.22}
\cleanpair{12.82}{8.68}{8.40}
\driftpair{13.18}{13.64}{12.22}
\cleanpair{13.82}{7.67}{6.28}
\driftpair{14.18}{13.96}{10.19}
\cleanpair{14.82}{24.58}{24.49}
\driftpair{15.18}{28.47}{27.96}
\cleanpair{15.82}{10.01}{7.93}
\driftpair{16.18}{14.83}{12.08}

% =========================================================
% (c) MAE - HiSTM-Nested
% =========================================================
\nextgroupplot[
    title={(c)},
    xmin=4, xmax=34,
    xtick={5,10,15,20,25,30},
    xlabel={MAE},
    yticklabels={}
]
\cleanpair{0.82}{14.06}{12.17}
\driftpair{1.18}{29.53}{22.84}
\cleanpair{1.82}{14.59}{10.61}
\driftpair{2.18}{29.07}{15.11}
\cleanpair{2.82}{8.18}{4.62}
\driftpair{3.18}{18.84}{8.12}
\cleanpair{3.82}{17.59}{16.83}
\driftpair{4.18}{23.60}{19.23}
\cleanpair{4.82}{11.84}{10.57}
\driftpair{5.18}{23.72}{16.81}
\cleanpair{5.82}{18.48}{8.32}
\driftpair{6.18}{32.12}{12.96}
\cleanpair{6.82}{22.21}{18.20}
\driftpair{7.18}{25.21}{19.33}
\cleanpair{7.82}{24.43}{24.10}
\driftpair{8.18}{28.31}{28.00}
\cleanpair{8.82}{16.63}{15.57}
\driftpair{9.18}{24.23}{18.85}
\cleanpair{9.82}{16.13}{11.95}
\driftpair{10.18}{28.82}{15.03}
\cleanpair{10.82}{20.33}{20.22}
\driftpair{11.18}{23.44}{21.00}
\cleanpair{11.82}{18.89}{14.22}
\driftpair{12.18}{23.01}{16.76}
\cleanpair{12.82}{6.89}{6.52}
\driftpair{13.18}{18.00}{16.66}
\cleanpair{13.82}{6.43}{4.95}
\driftpair{14.18}{22.21}{11.27}
\cleanpair{14.82}{18.57}{16.98}
\driftpair{15.18}{24.70}{20.48}
\cleanpair{15.82}{8.41}{7.24}
\driftpair{16.18}{22.39}{12.97}

% =========================================================
% (d) RMSE - HiSTM-Nested
% =========================================================
\nextgroupplot[
    title={(d)},
    xmin=6, xmax=39,
    xtick={10,15,20,25,30,35},
    xlabel={RMSE},
    yticklabels={}
]
\cleanpair{0.82}{18.30}{16.08}
\driftpair{1.18}{33.68}{26.80}
\cleanpair{1.82}{19.33}{14.16}
\driftpair{2.18}{32.94}{18.68}
\cleanpair{2.82}{11.49}{6.51}
\driftpair{3.18}{21.58}{10.25}
\cleanpair{3.82}{24.33}{23.21}
\driftpair{4.18}{29.46}{25.38}
\cleanpair{4.82}{15.29}{13.64}
\driftpair{5.18}{27.21}{20.22}
\cleanpair{5.82}{25.02}{11.75}
\driftpair{6.18}{37.16}{15.76}
\cleanpair{6.82}{28.77}{23.78}
\driftpair{7.18}{31.25}{24.59}
\cleanpair{7.82}{33.04}{32.59}
\driftpair{8.18}{37.47}{37.05}
\cleanpair{8.82}{22.52}{21.16}
\driftpair{9.18}{29.36}{24.10}
\cleanpair{9.82}{25.37}{19.27}
\driftpair{10.18}{34.99}{21.27}
\cleanpair{10.82}{26.12}{25.99}
\driftpair{11.18}{29.24}{26.70}
\cleanpair{11.82}{26.10}{19.60}
\driftpair{12.18}{29.07}{21.41}
\cleanpair{12.82}{9.07}{8.59}
\driftpair{13.18}{20.11}{18.66}
\cleanpair{13.82}{9.06}{7.20}
\driftpair{14.18}{23.94}{14.07}
\cleanpair{14.82}{25.18}{23.03}
\driftpair{15.18}{30.83}{26.24}
\cleanpair{15.82}{11.25}{9.69}
\driftpair{16.18}{25.03}{16.10}

\end{groupplot}

% Manual legend
\node[
    anchor=north,
    font=\footnotesize,
    inner sep=0pt
] at ($(group c2r1.south east)!0.5!(group c3r1.south west)+(0,-0.85cm)$) {
    \begin{tikzpicture}
        \draw[CleanBlue!55, line width=0.95pt] (0,0) -- (0.35,0);
        \draw[
            mark=*,
            mark size=2.05pt,
            mark options={fill=white, draw=CleanBlue, line width=0.65pt}
        ] plot coordinates {(0,0)};
        \node[anchor=west, font=\footnotesize] at (0.45,0) {BL};

        \draw[CleanBlue!55, line width=0.95pt] (1.25,0) -- (1.60,0);
        \draw[
            mark=*,
            mark size=2.05pt,
            mark options={fill=CleanBlue, draw=CleanBlue, line width=0.65pt}
        ] plot coordinates {(1.25,0)};
        \node[anchor=west, font=\footnotesize] at (1.70,0) {BL+PID};

        \draw[DriftOrange!65, line width=1.00pt] (2.85,0) -- (3.20,0);
        \draw[
            mark=square*,
            mark size=2.15pt,
            mark options={fill=white, draw=DriftOrange, line width=0.65pt}
        ] plot coordinates {(2.85,0)};
        \node[anchor=west, font=\footnotesize] at (3.30,0) {BL+D};

        \draw[DriftOrange!65, line width=1.00pt] (4.20,0) -- (4.55,0);
        \draw[
            mark=square*,
            mark size=2.15pt,
            mark options={fill=DriftOrange, draw=DriftOrange, line width=0.65pt}
        ] plot coordinates {(4.20,0)};
        \node[anchor=west, font=\footnotesize] at (4.65,0) {BL+D+PID};
    \end{tikzpicture}
};

\end{tikzpicture}

\caption{Per-cell error under the \textit{Hotspot Linear Local (HLL)} drift scenario for the selected cells. Panels (a) and (b) report MAE and RMSE, respectively, for HiSTM, while panels (c) and (d) report MAE and RMSE, respectively, for HiSTM-Nested. For each cell, the upper dumbbell shows the clean baseline pair, BL to BL+PID, and the lower dumbbell shows the drifted pair, BL+D to BL+D+PID. A leftward displacement indicates reduced error after PID correction.}
\label{fig:percell_dumbbell_bl_pid_drift}
\end{figure*}
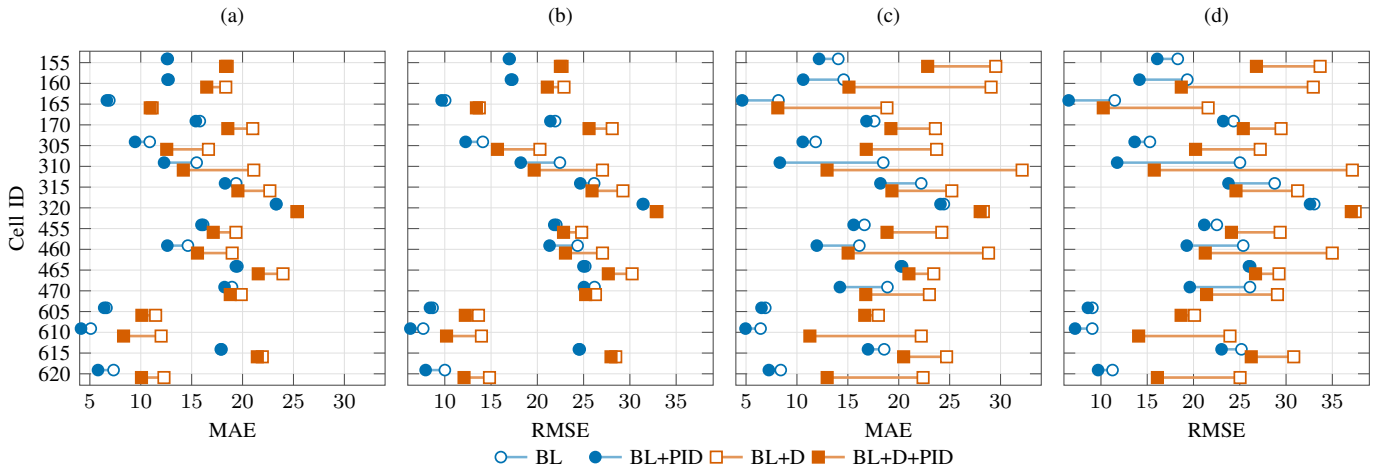

% \begin{figure}[]
% \centering

% \begin{subfigure}{\columnwidth}
%     \centering
%     \includegraphics[width=\columnwidth]{images/per_cell_mae.png}
%     \caption{Per cell MAE}
%     \label{fig:percell_mae}
% \end{subfigure}

% \vspace{0.5cm}

% \begin{subfigure}{\columnwidth}
%     \centering
%     \includegraphics[width=\columnwidth]{images/per_cell_rmse.png}
%     \caption{Per cell RMSE}
%     \label{fig:percell_rmse}
% \end{subfigure}
% \end{figure}

To provide a per-cell analysis on 16 selected out cells, and due to page limit constrain, a representative 
subset of figures is presented here, with the remaining results following similar and consistent 
trends. Figures~\ref{fig:mitigation_heatmaps} (a and b) and \ref{fig:avg_mitigation} 
 present cell-level MAE and RMSE percentage mitigation results under PID correction across all drift 
scenarios. Both heatmaps reveal consistent drift mitigation across all drift 
conditions, with the most pronounced improvements observed under the scenario
\textit{Hotspot\_Linear\_Local}, which exhibits the highest mitigation percentage 
across the majority of cells. Even the more spatially distributed 
\textit{Joint\_ST} scenarios displayed significant mitigation percentages of 
$\Delta\text{MAE},\, \Delta\text{RMSE} \gtrsim 10\%$ across most cells, confirming 
that the PID correction remains effective regardless of the performance metric considered.

Figure~\ref{fig:avg_mitigation} further consolidates these findings through 
the average mitigation summary aggregated across the 16 selected out cells. 
The \textit{Hotspot\_Linear\_Local} scenario yields the highest average mitigation, 
reaching $\overline{\Delta\text{MAE}} \approx 30.18\%$ and $\overline{\Delta\text{RMSE}} 
\approx 26.68\%$ for \textit{HiSTM\_Nested} model, and $\approx 12.74\%$ and $\approx 10.90\%$ 
for \textit{HiSTM} model. The remaining drift scenarios exhibit lower but 
sustained average mitigation values of $\overline{\Delta\text{MAE}} \in [17\%, 26\%]$ 
for \textit{HiSTM\_Nested} and $\overline{\Delta\text{MAE}} \in [6\%, 9\%]$ for 
\textit{HiSTM}, confirming that PID 
correction is beneficial across all scenarios. Additionally, Figure~\ref{fig:percell_dumbbell_bl_pid_drift} presents the per-cell MAE and RMSE under the \textit{Hotspot Linear Local (HLL)} drift scenario for both \textit{HiSTM} (panels a--b) and \textit{HiSTM\_Nested} (panels c--d). Across nearly all cells, the lower dumbbell markers exhibit a clear leftward displacement from BL+D to BL+D+PID, confirming that PID correction systematically reduces drift-induced prediction error. In several cells, this recovery is substantial, approaching the clean-data performance captured by the upper BL-to-BL+PID dumbbell, thus highlights the robustness of the PID mechanism .

\section{conclusion}
\label{sec:conclusion}
Our work demonstrated the effectiveness of integrating a PID-based correction framework for spatiotemporal traffic forecasting, operating entirely online at inference time to deliver lightweight low-cost adjustments without model retraining, making it well-suited for real-time deployment in dynamic network environments. Results confirm consistent improvements in prediction accuracy across diverse traffic drift scenarios, highlighting its adaptability to the conditions expected in future 6G networks. Nevertheless, future research directions will focus on investigating the sensitivity of the PID controller to its control parameters, the impact of tuning strategies, and the influence of the underlying baseline model accuracy on overall performance. Additionally, extension of the formwork to more advanced forecasting architectures will be considered.

\section*{Acknowledgment}
Parts of this work has been funded by the Bavarian Government through the High-Tech Agenda (HTA).

\bibliographystyle{unsrt}
\bibliography{references}
\end{document}